\documentclass[apj]{emulateapj}
\usepackage{graphics}
\usepackage{subfigure}
\usepackage{epsfig}
\usepackage{alltt}
\usepackage{verbatim}

\def\spose#1{\hbox to 0pt{#1\hss}}
\def\simlt{\mathrel{\spose{\lower 3pt\hbox{$\mathchar"218$}}
    \raise 2.0pt\hbox{$\mathchar"13C$}}}
\def\simgt{\mathrel{\spose{\lower 3pt\hbox{$\mathchar"218$}}
    \raise 2.0pt\hbox{$\mathchar"13E$}}}

\newcommand{\oiii}{\mbox{[\ion{O}{3}]} $\,$}
\newcommand{\oiiiw}{\mbox{[\ion{O}{3}] $\lambda$5007} $\,$}
\newcommand{\oiiiwn}{\mbox{[\ion{O}{3}] $\lambda$5007}}

\newcommand{\oiiiwwn}{\mbox{[\ion{O}{3}] $\lambda$4959}}
\newcommand{\hb}{\mbox{H$\beta$} $\,$}
\newcommand{\hbn}{\mbox{H$\beta$}}

\shortauthors{Comerford et al.}
\shorttitle{Kiloparsec-scale Spatial Offsets in Double-peaked Narrow-line Active Galactic Nuclei}

\begin{document}

\title{Kiloparsec-scale Spatial Offsets in Double-peaked Narrow-line Active Galactic Nuclei. I. Markers for Selection of Compelling Dual \\ Active Galactic Nuclei Candidates}

\author{Julia M. Comerford\altaffilmark{1,9}, Brian F. Gerke\altaffilmark{2,10}, Daniel Stern\altaffilmark{3}, Michael C. Cooper\altaffilmark{4,11}, Benjamin J. Weiner\altaffilmark{5}, \\ Jeffrey A. Newman\altaffilmark{6}, Kristin Madsen\altaffilmark{7}, and R. Scott Barrows\altaffilmark{8}}

\thanks{$^9$NSF Astronomy and Astrophysics Postdoctoral Fellow}
\thanks{$^{10}$Current address: Lawrence Berkeley National Laboratory, 1 Cyclotron Road, MS 90-4000, Berkeley, CA 94720, USA}
\thanks{$^{11}$Hubble Fellow}

\affil{$^1$Astronomy Department, University of Texas at Austin, Austin, TX 78712, USA}
\affil{$^2$Kavli Institute for Particle Astrophysics and Cosmology,
  M/S 29, Stanford Linear Accelerator Center, \\ 2575 Sand Hill Road, 
  Menlo Park, CA 94725, USA}
\affil{$^3$Jet Propulsion Laboratory, California Institute of Technology, MS 169-221, 4800 Oak Grove Drive, Pasadena, CA 91109, USA}
\affil{$^4$Center for Galaxy Evolution, Department of Physics and Astronomy, University of California, Irvine, 4129 Frederick Reines Hall, Irvine, CA 92697, USA}
\affil{$^5$Steward Observatory, University of Arizona, 933 North Cherry Avenue, Tucson, AZ 85721, USA}
\affil{$^6$Pittsburgh Particle Physics, Astrophysics, and Cosmology Center, Department of Physics and Astronomy, University of Pittsburgh, Pittsburgh, PA 15260, USA}
\affil{$^7$Space Radiation Laboratory, California Institute of Technology, MS 105-24, Pasadena, CA 91125, USA}
\affil{$^8$Arkansas Center for Space and Planetary Sciences, University of Arkansas, Fayetteville, AR 72701, USA}

\begin{abstract}
Merger-remnant galaxies with kiloparsec (kpc) scale separation dual active galactic nuclei (AGNs) should be widespread as a consequence of galaxy mergers and triggered gas accretion onto supermassive black holes, yet very few dual AGNs have been observed.  Galaxies with double-peaked narrow AGN emission lines in the Sloan Digital Sky Survey are plausible dual AGN candidates, but their double-peaked profiles could also be the result of gas kinematics or AGN-driven outflows and jets on small or large scales.  To help distinguish between these scenarios, we have obtained spatial profiles of the AGN emission via follow-up long-slit spectroscopy of 81 double-peaked narrow-line AGNs in SDSS at $0.03 \leq z \leq 0.36$ using Lick, Palomar, and MMT Observatories.  We find that all 81 systems exhibit double AGN emission components with $\sim$kpc projected spatial separations on the sky (0.2 $h^{-1}_{70}$ kpc $< \Delta x <$ 5.5 $h^{-1}_{70}$ kpc; median $\Delta x=1.1$ $h^{-1}_{70}$ kpc), which suggests that they are produced by kpc-scale dual AGNs or kpc-scale outflows, jets, or rotating gaseous disks.  Further, the objects split into two subpopulations based on the spatial extent of the double emission components and the correlation between projected spatial separations and line-of-sight velocity separations. These results suggest that the subsample ($58^{+5}_{-6} \, \%$) of the objects with spatially compact emission components may be preferentially produced by dual AGNs, while the subsample ($42^{+6}_{-5} \,\%$) with spatially extended emission components may be preferentially produced by AGN outflows.  We also find that for $32^{+8}_{-6} \, \%$ of the sample the two AGN emission components are preferentially aligned with the host galaxy major axis, as expected for dual AGNs orbiting in the host galaxy potential.  Our results both narrow the list of possible physical mechanisms producing the double AGN components, and suggest several observational criteria for selecting the most promising dual AGN candidates from the full sample of double-peaked narrow-line AGNs.  Using these criteria, we determine the 17 most compelling dual AGN candidates in our sample.
\end{abstract}  


\section{Introduction}
\label{intro}

It is now established that most bulge-dominated galaxies host central supermassive black holes (SMBHs) and that galaxies frequently merge with one another.  As a natural consequence, there must exist a subpopulation of galaxies that harbor two SMBHs. These SMBH pairs are observable if they are accreting gas as active galactic nuclei (AGNs), though some may be enshrouded by dust.  The kiloparsec (kpc) scale separation dual AGNs are an intermediate evolutionary stage between the galaxy pairs separated by tens of kpc used as proxies for the merger rate (e.g., \citealt{LI08.1,EL08.1,DE11.1}) and the sub-pc separation binary SMBHs (e.g., \citealt{ER11.1}) that are expected to produce gravitational waves upon coalescence.  Although there are thousands of known galaxy pairs with tens of kpc separations, only a handful of kpc-scale dual AGNs \citep{KO03.1,HU06.1,BI08.1,KO11.1,FU11.3}, one 100 pc scale AGN pair \citep{FA11.1}, and one 10 pc scale AGN pair \citep{RO06.1} have been confirmed.

The small number of confirmed dual AGNs is in conflict with expectations based on the galaxy merger rate and the triggering of gas accretion onto SMBHs during mergers (e.g., \citealt{SP05.1,HO05.1,VA12.1}).  Although discoveries of individual dual AGN systems continue (e.g., \citealt{CO09.3,BA12.1}), a systematic survey of dual AGNs is now necessary to close the gap between the observed and expected numbers.

Such a systematic survey of dual AGNs would shed light on drivers of galaxy evolution such as galaxy mergers and SMBH growth.  As direct observational tracers of galaxy mergers, dual AGNs can be used to estimate the galaxy merger rate.  Further, the number of dual SMBHs powering AGNs can set constraints on the activation probability of SMBHs in merging galaxies, which is a key parameter in models of the growth of SMBHs by gas accretion during galaxy mergers.

With the advent of large spectroscopic surveys of galaxies, it is now possible to conduct a systematic census of dual AGNs.  Two dual AGN candidates were identified in the DEEP2 Galaxy Redshift Survey based on their long-slit spectra, which revealed two components of AGN-fueled \oiiiw emission separated by $\sim$kpc spatially on the sky and a few hundred km s$^{-1}$ in line-of-sight velocity \citep{GE07.2,CO09.1}.  The velocity separations manifest as double-peaked emission lines in the one-dimensional spectra.  However, the extreme rarity of double-peaked AGNs demands a larger survey for statistical studies.

Recently, 340 unique AGNs with double-peaked \oiiiw were identified in the Sloan Digital Sky Survey (SDSS) at $0.01<z<0.7$ \citep{WA09.1, LI10.1, SM10.1}.  These are plausible dual AGN candidates, but velocity offsets in AGN emission lines can also be produced by narrow-line region (NLR) effects such as small-scale ($\simlt$ 100 pc) gas kinematics and outflows (e.g., \citealt{DA05.2,DA06.1,SM12.1}) or larger-scale AGN outflows (e.g., \citealt{CR10.1, FI11.1}).  Since SDSS fiber spectra carry insufficient spatial information, spatially resolved follow-up slit spectroscopy is required to determine both whether the double-peaked emission lines come from spatially distinct regions and what the physical scales of the regions are.  These observations will help constrain which physical mechanisms are producing the double-peaked AGN lines and which objects are good candidates for dual AGNs.

We have obtained follow-up long-slit observations of 81 (or one-fourth) of the double-peaked narrow-line AGNs in SDSS using a combination of the Kast Spectrograph at Lick Observatory, the Double Spectrograph at Palomar Observatory, and the Blue Channel Spectrograph at MMT Observatory.  Since the orientation of the double emission components on the sky is unknown, we observe each object at two position angles, generally separated by 90$^\circ$, so that we can determine the full spatial separation of the two emission components on the sky as well as their position angle.  We use this information, as well as the spatial extent of the emission and existing multiwavelength observations of the host galaxies, to constrain what mechanisms (gas rotation, AGN outflows and jets, or dual AGNs) on what scales ($\simlt100$ pc, $\sim$kpc, or $\sim$10 kpc) produce the line profiles, and to identify which objects are the best candidates for dual AGNs.

We assume a Hubble constant $H_0 =70$ km s$^{-1}$ Mpc$^{-1}$, $\Omega_m=0.3$, and $\Omega_\Lambda=0.7$ throughout, and all distances are given in physical (not comoving) units.

\section{The Sample and Observations}

\newpage

\begin{deluxetable*}{llllll}
\tablewidth{0pt}
\tablecolumns{6}
\tablecaption{Summary of Observations} 
\tablehead{
\colhead{SDSS Designation} &
\colhead{Telescope/Instrument} & 
\colhead{Observation Date (UT)} &
\colhead{$\theta_{1}$ ($^\circ$)} &
\colhead{$\theta_{2}$ ($^\circ$)} &
\colhead{Exposure Time (s)} 
}
\startdata 
SDSS J000249.07+004504.8 & Lick/Kast & 2009 August 17 / 2009 August 18 & 69.4 & 158.0 & 3600 \\
SDSS J000656.85+154847.9 & MMT/Blue Channel & 2010 November 6 & 44.4 & 134.4 & 1080 \\
SDSS  J000911.58$-$003654.7 & MMT/Blue Channel & 2010 November 6 & 67.0 & 157.0 & 1080 \\
SDSS J011659.59$-$102539.1 & MMT/Blue Channel & 2010 November 5 & 28.8 & 118.8 & 1080 \\
\enddata
\tablecomments{We observed each object at two position angles, $\theta_{1}$ and $\theta_{2}$ (given in degrees east of north), and the exposure time given is for each position angle. \\ (This table is available in its entirety in a machine-readable form in the online journal.  A portion is shown here for guidance regarding its form and content.)}
\label{tbl-1}
\end{deluxetable*}

\begin{deluxetable*}{ccclcccccccc}
\tablewidth{0pt}
\tablecolumns{12}
\tablecaption{Summary of Measurements} 
\tablehead{
\colhead{SDSS} &
\colhead{Redshift} & 
\colhead{Spec.} &
\colhead{$\Delta v$} &
\colhead{$\Delta x$} &
\colhead{$\Delta x$} &
\colhead{Phys.} &
\colhead{$\theta_{\rm sky}$} &
\colhead{$\Delta\theta$} & 
\colhead{$e$} &
\colhead{Other Obs.} &
\colhead{Reference} \\
 Name & & Type & (km s$^{-1}$) & ($^{\prime\prime}$) & ($h^{-1}_{70}$ kpc) & Extent & ($^\circ$) & ($^\circ$) & & &
}
\startdata 
0002+0045 & 0.087 & 2 & 511 $\pm$ 3 & \footnotesize{0.58 $\pm$ 0.05} & 0.95 $\pm$ 0.08 & c & \footnotesize{\phantom{0}63.9 $\pm$ 5.0} & 46.3 & 0.106 & NIR/SS/VLBA & 1, 2, 3, 4, 5 \\
0006+1548 & 0.125 & 2 & 359 $\pm$ 3 & \footnotesize{0.37 $\pm$ 0.02} & 0.84 $\pm$ 0.05 & e & \footnotesize{172.6 $\pm$ 3.9} & 51.8 & 0.040 & NIR & 2, 3 \\
0009$-$0036 & 0.073 & 2 & 304 $\pm$ 5 & \footnotesize{0.25 $\pm$ 0.01} & 0.35 $\pm$ 0.01 & c & \footnotesize{\phantom{0}56.2 $\pm$ 1.1} & 79.1 & 0.188 & NIR/SS & 1, 4 \\
0116$-$1025 & 0.150 & 2 & 288 $\pm$ 5 & \footnotesize{1.02 $\pm$ 0.06} & 2.68 $\pm$ 0.16 & c & \footnotesize{117.0 $\pm$ 2.6} & \phantom{0}1.8 & 0.151 & NIR/SS & 1, 4 \\
\enddata
\tablecomments{Column 3 shows the spectral type of the AGNs.  Column 4 shows the velocity offset between the \oiiiw peaks in the SDSS spectrum.  Column 5 shows the angular projected spatial offset between the two \oiiiw emission features on the sky, as measured from our slit spectroscopy.  Column 6 shows the physical projected spatial offset.  Column 7 shows whether the double AGN emission features in the two-dimensional spectrum appear compact (c) or extended (e) in physical extent.  Column 8 shows the position angle between the two \oiiiw emission features on the sky, measured from our slit spectroscopy in degrees east of north.  Column 9 shows the absolute value of the difference between the measured position angle and the isophotal position angle of the major axis of the object from SDSS $r$-band photometry (with the same error as in Column 8). Column 10 shows the ellipticity of the object from SDSS photometry.  Column 11 lists other published observations of the object: near infrared imaging (NIR), slit spectroscopy (SS), Very Long Baseline Array observations (VLBA), integral field unit spectroscopy (IFU), {\it Hubble Space Telescope} imaging ({\it HST}), Very Large Array observations (VLA), and {\it Chandra} observations ({\it Chandra}). References: (1) \cite{LI10.1}, (2) \cite{WA09.1}, (3) \cite{FU12.1}, (4) \cite{SH11.1}, (5) \cite{TI11.1}, (6) \cite{SM10.1},  (7) \cite{RO11.1}, (8) \cite{FU11.1}, (9) \cite{MC11.1}, (10) \cite{RO10.1}, (11) \cite{CO11.2}. \\ (This table is available in its entirety in a machine-readable form in the online journal.  A portion is shown here for guidance regarding its form and content.)}
\label{tbl-2}
\end{deluxetable*}

\subsection{The Sample}

We selected targets from the three catalogs that have identified active galaxies in SDSS with double-peaked \oiiiw emission lines: \cite{WA09.1}, \cite{LI10.1}, and \cite{SM10.1} (we note that one of the objects in \citealt{SM10.1} was originally identified in \citealt{XU09.1}).  These AGNs were selected using standard emission line diagnostics \citep{BA81.1,KE01.2}, and the double-peaked AGN were selected with varying techniques.  \cite{WA09.1} select 87 double-peaked Type 2 AGNs that have blueshifted and redshifted components of \oiiiwn, relative to the host galaxy redshifts, that have a wavelength difference $\ge 1 \, \mathrm{\AA}$ and a flux ratio between the two components of 0.3 -- 3; \cite{LI10.1} select 167 double-peaked Type 2 AGNs by eye, through visual identification of systems that have similar double-peaked profiles in both \oiiiw and \oiiiwwn; \cite{SM10.1} select 86 double-peaked Type 1 AGNs and 62 double-peaked Type 2 AGNs through visual selection of systems with double peaks in both \oiiiw and \oiiiwwn.  After accounting for the overlaps between these samples, there are 340 unique double-peaked AGNs. We drew our targets from these 340 objects.  

Since our aim was to use follow-up long-slit spectroscopy of these objects to measure $\sim$kpc spatial offsets between the \oiiiw emission components, we selected targets as double-peaked AGNs that had 1) airmasses below 1.5 during our observing runs, 2) redshifts $z \leq 0.36$ where $\simlt$kpc spatial offsets were resolvable with the spectrograph pixel scale, and 3) sufficiently bright $r$-band magnitudes $r \leq 18.5$ to minimize exposure times and maximize the number of objects observed in this campaign.  The latter two criteria are spectrograph- and telescope-dependent, and we selected targets for each telescope based on the spectrograph pixel size and the aperture of the telescope as described below.  We made no selections on any other galaxy properties.

We obtained follow-up slit spectroscopy for a sample of 81 double-peaked AGNs at $0.03 \leq z \leq 0.36$ and $14.5 \leq r \leq 18.5$.  These targets are a representative sample of the 237 double-peaked AGN in SDSS that fit the criteria $z \leq 0.36$ and $r \leq 18.5$; Kolmogorov-Smirnov tests find a 91$\%$ (87$\%$) probability that the redshifts ($r$-band magnitudes) of our observed sample and the 237 potential targets were derived from the same parent distribution.  We note that this is a pilot program, and long-slit observations of more double-peaked AGN in SDSS are forthcoming (J. M. Comerford et al., in preparation).

\subsection{Observations}

We obtained long-slit spectroscopy of the 81 double-peaked AGNs using the 
Kast Spectrograph at the Lick 3 m telescope (pixel size $0\farcs78$), the Double Spectrograph (DBSP) on the Palomar 5 m telescope (pixel size $0\farcs47$ for the red detector and $0\farcs39$ for the blue detector), and the Blue Channel Spectrograph on the MMT 6.5 m telescope (pixel size $0\farcs29$). We optimized the aperture size and pixel size of each telescope-spectrograph combination to observe the lower redshift and brighter objects with Lick/Kast, the medium redshift and brightness objects with Palomar/DBSP, and the higher redshift and fainter objects with MMT/Blue Channel. We note that since we resolve each double-peaked AGN's two emission components in velocity space, it is possible to measure their positional centroids to much better than the seeing limit and that the limiting angular scale is instead the pixel scale.

We observed subsamples of our target list as follows: 16 objects at $0.04 < z < 0.16$ (median $z=0.08$, where 1$^{\prime\prime}$ corresponds to 1.5 $h_{70}^{-1}$ kpc) and $14.5 < r < 17.5$ with Lick/Kast; 16 objects at $0.06 < z < 0.15$ (median $z=0.11$, where 1$^{\prime\prime}$ corresponds to 2.0 $h_{70}^{-1}$ kpc) and $15.0 < r < 18.2$ with Palomar/DBSP; and 49 objects at $0.03 < z < 0.36$ (median $z=0.15$, where 1$^{\prime\prime}$ corresponds to 2.6 $h_{70}^{-1}$ kpc) and $14.5 < r < 18.5$ with MMT/Blue Channel. Table~\ref{tbl-1} summarizes the observations.  We used a 1200 lines mm$^{-1}$ grating with each telescope/spectrograph, centered such that the wavelength range spanned \hb and \oiii for the redshift range of each night's subsample of targets. 

We observed each target twice, with the slit at two different position angles, in order to determine the orientation of the AGN emission components on the plane of the sky.  We typically made observations at the isophotal position angle of the major axis of the object in SDSS $r$-band photometry and the corresponding orthogonal position angle.  However, if the SDSS imaging showed a companion near our target we aligned the position angle to include that companion rather than using the position angle of the galaxy, and at times mechanical constraints on the rotation of the telescope prevented us from observing at a second position angle that was orthogonal to the first.

The data were reduced following standard procedures in IRAF and IDL.

\section{Analysis of SDSS Data}

\subsection{Velocity Measurements}

In this section we present measurements of the velocity offsets between the redshifted and blueshifted components of the double-peaked \oiiiw emission lines in the SDSS spectra.  Although velocity offsets have been measured in the literature \citep{WA09.1, LI10.1, SM10.1}, cases where different authors measured the velocity offset for the same object often conflict.  Consequently, for completeness and consistency, we measure all of the velocity offsets from the SDSS spectra here (Table~\ref{tbl-2}).

For each galaxy, we fit two Gaussians to the continuum-subtracted \oiiiw emission line profile, which is the emission line with the highest signal-to-noise ratio in our observations.  We measure the line-of-sight velocity difference based on the wavelengths of the peaks of the best-fit Guassians, and the error in the velocity difference is derived from the errors in the peak wavelengths of the best-fit Guassians added in quadrature.

\subsection{Ellipticities}
\label{ellipticities}
\begin{figure}
\centering
\subfigure{\includegraphics[height=4.6cm]{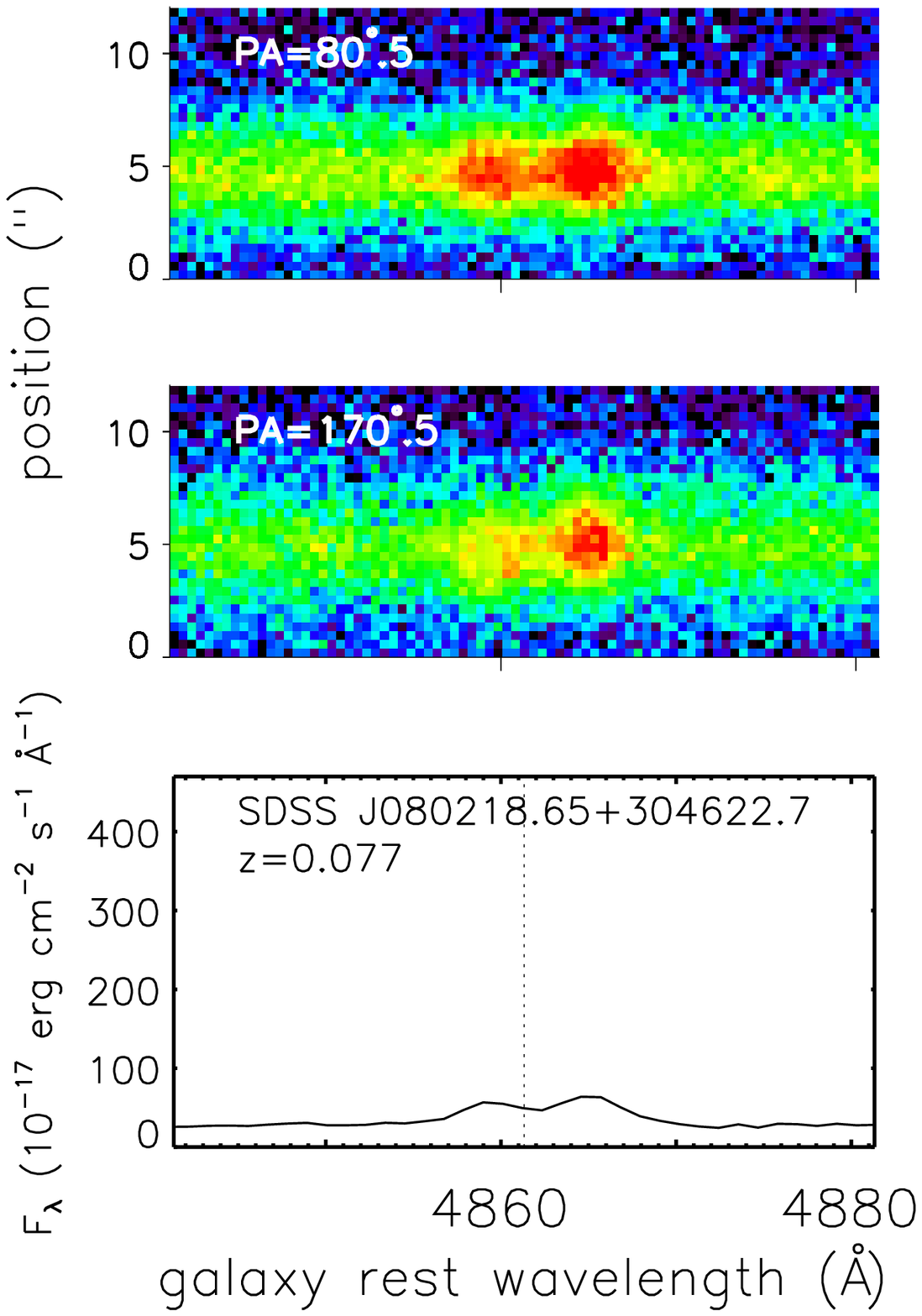}}
\hspace{-0.7cm}
\subfigure{\includegraphics[height=4.6cm]{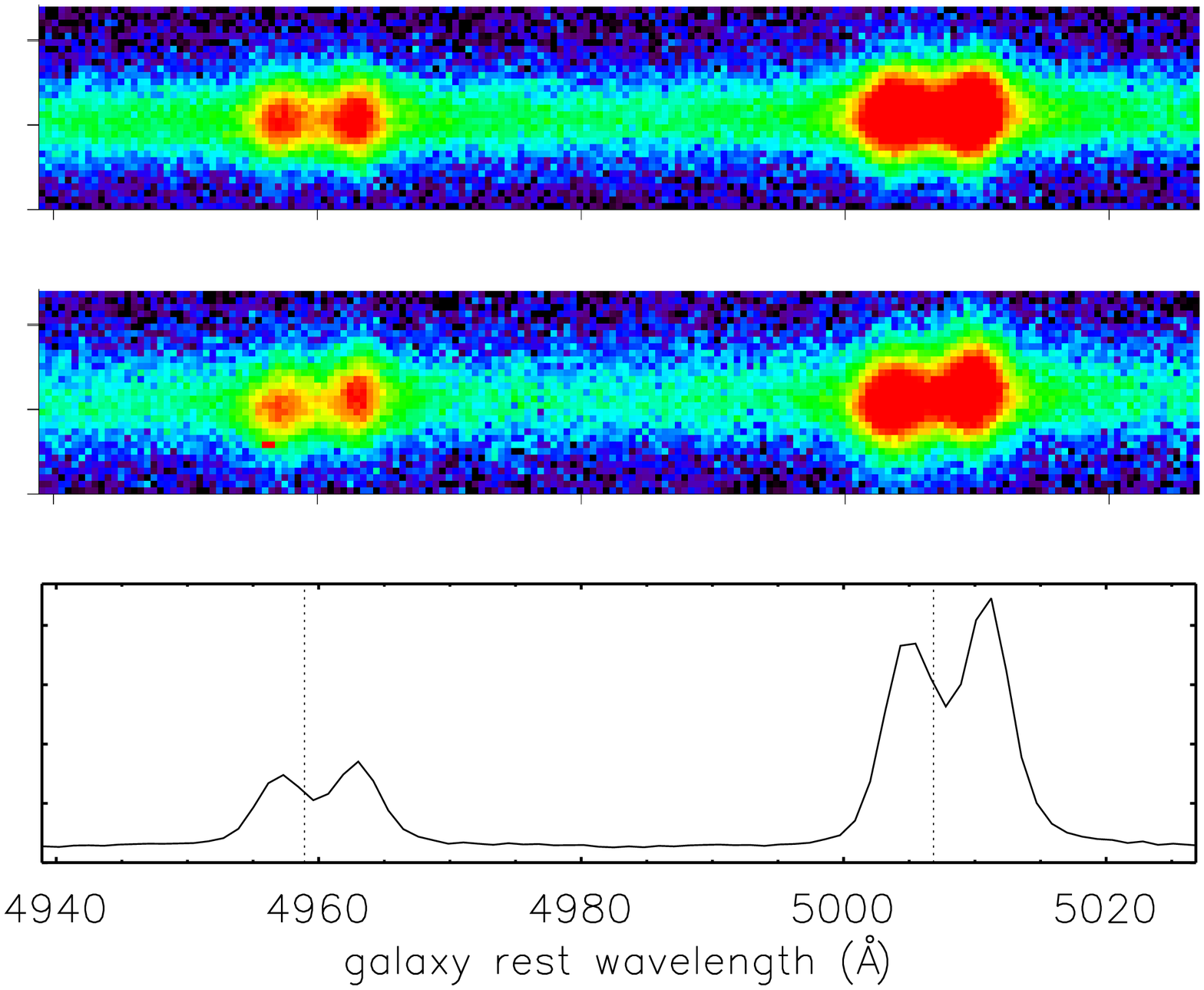}} 
\subfigure{\includegraphics[height=3.9cm]{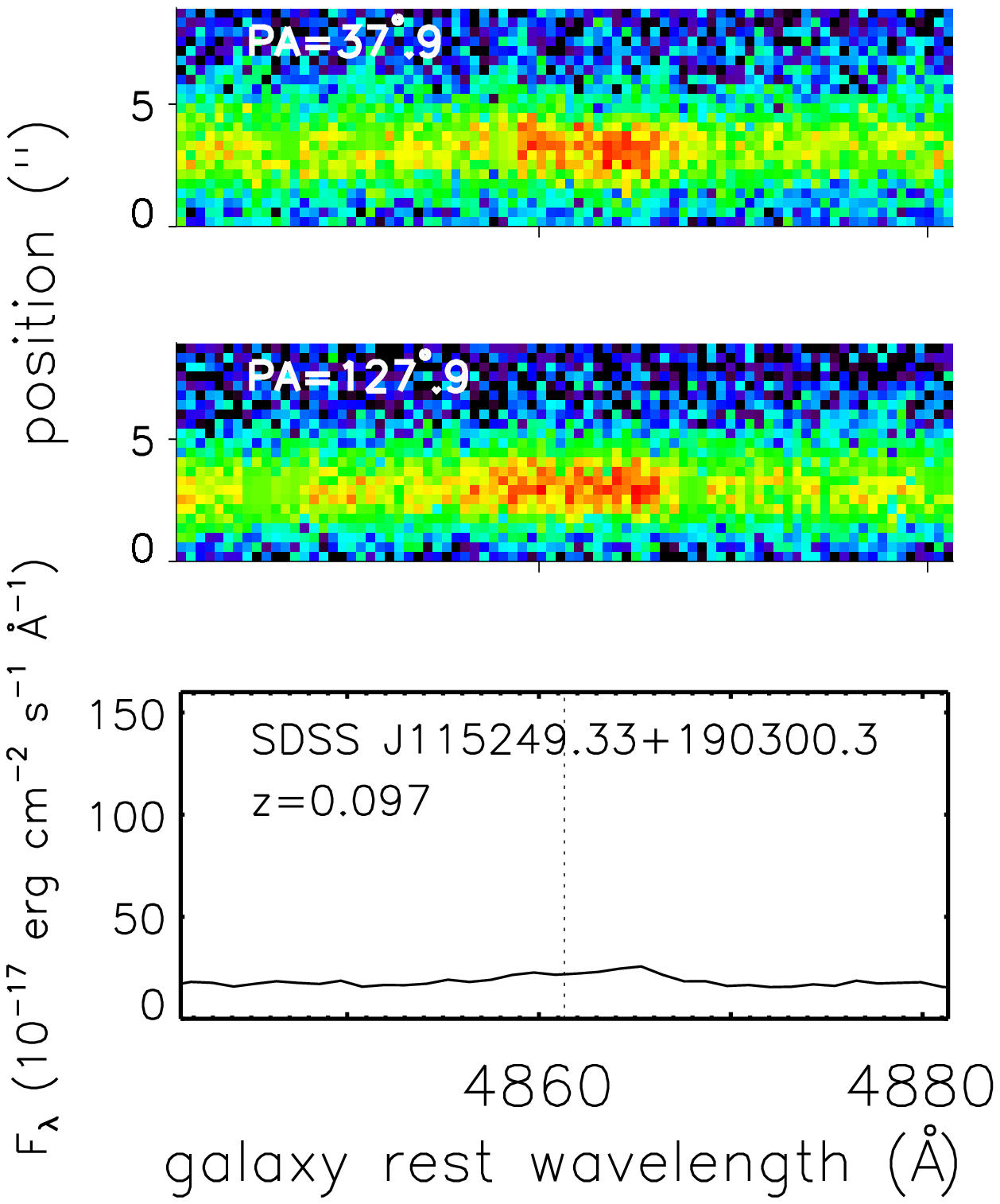}}
\hspace{-0.7cm}
\subfigure{\includegraphics[height=3.9cm]{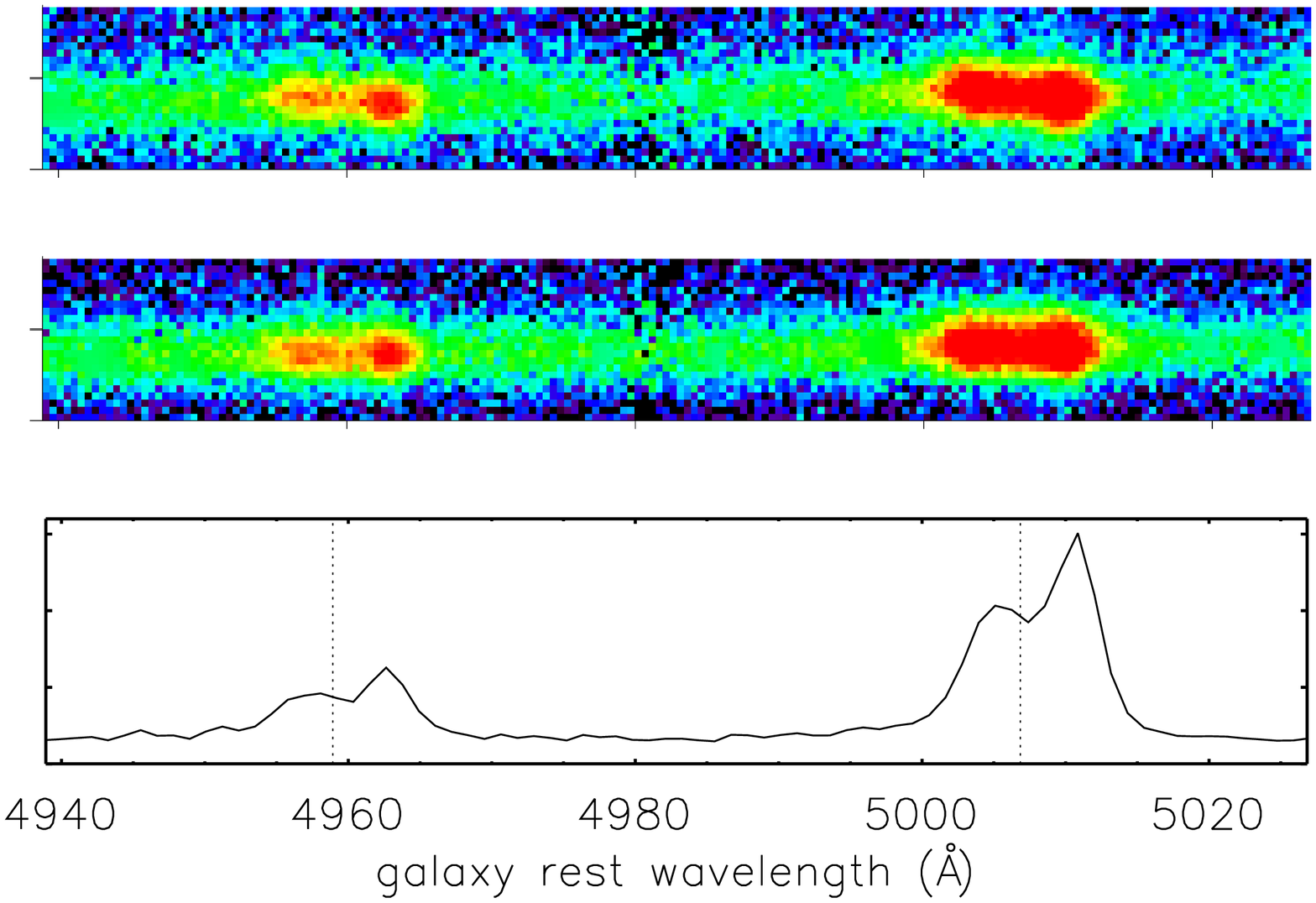}} 
\subfigure{\includegraphics[height=4.4cm]{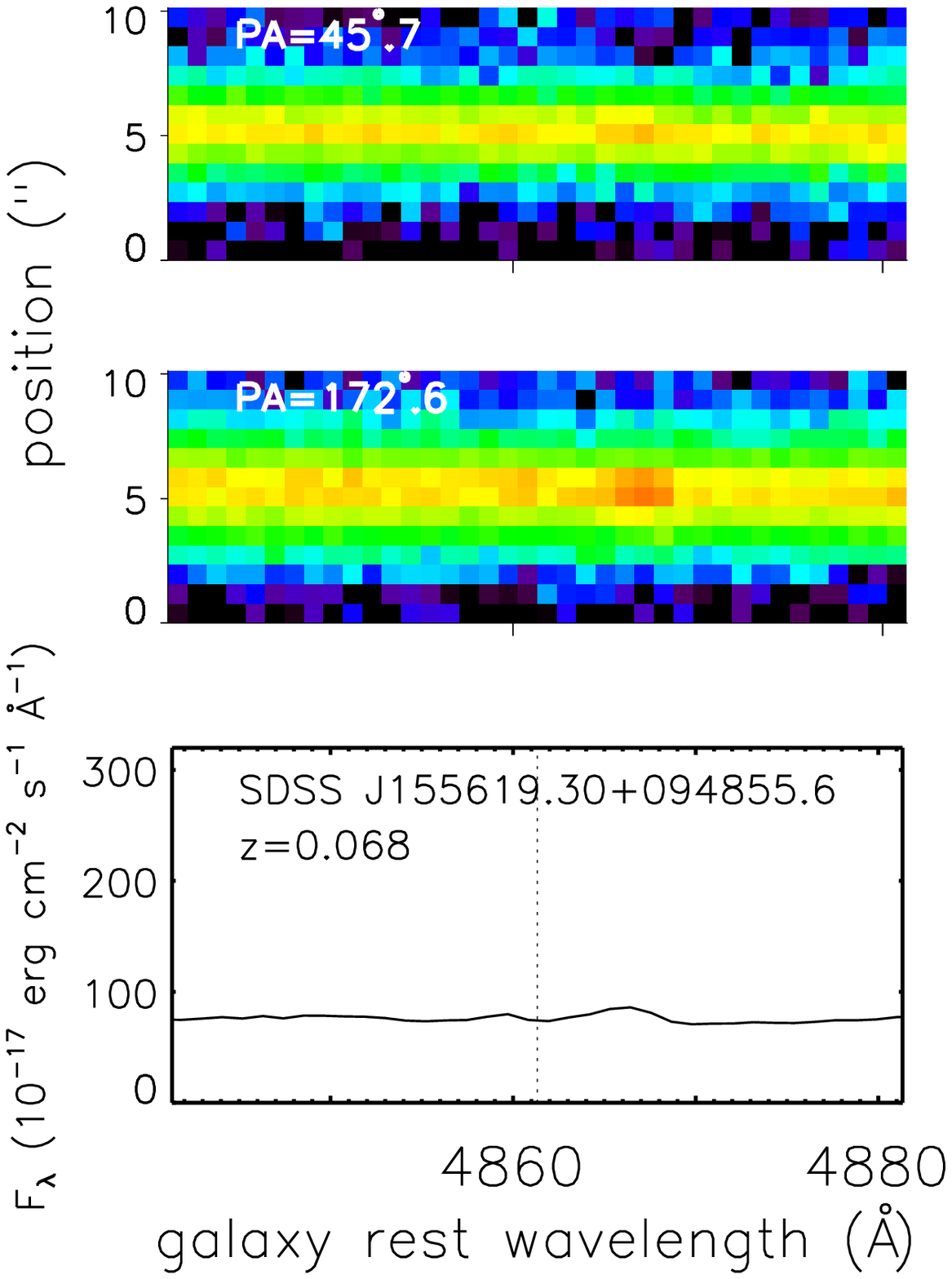}}
\hspace{-0.7cm}
\subfigure{\includegraphics[height=4.4cm]{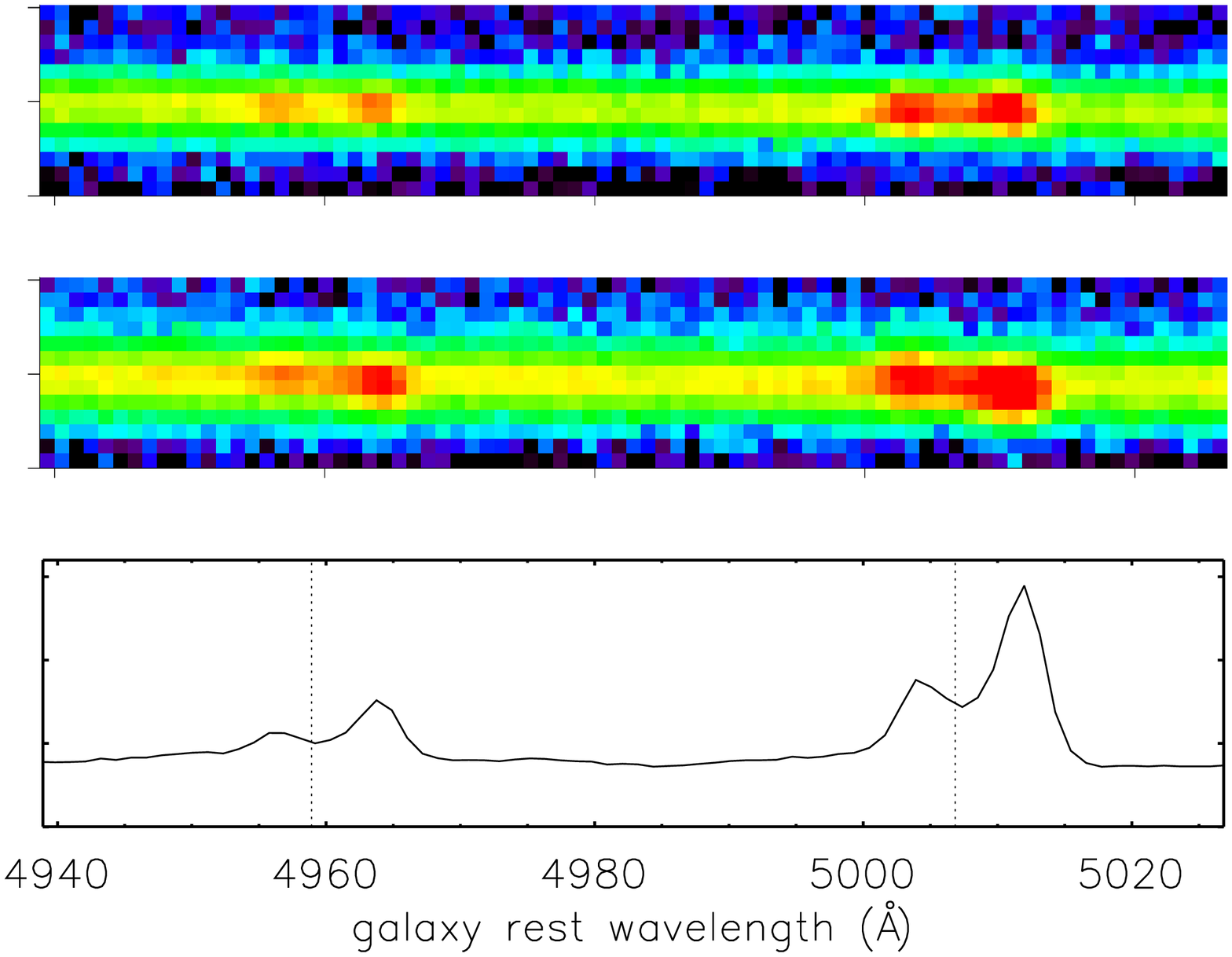}} 
\subfigure{ \includegraphics[height=3.44cm]{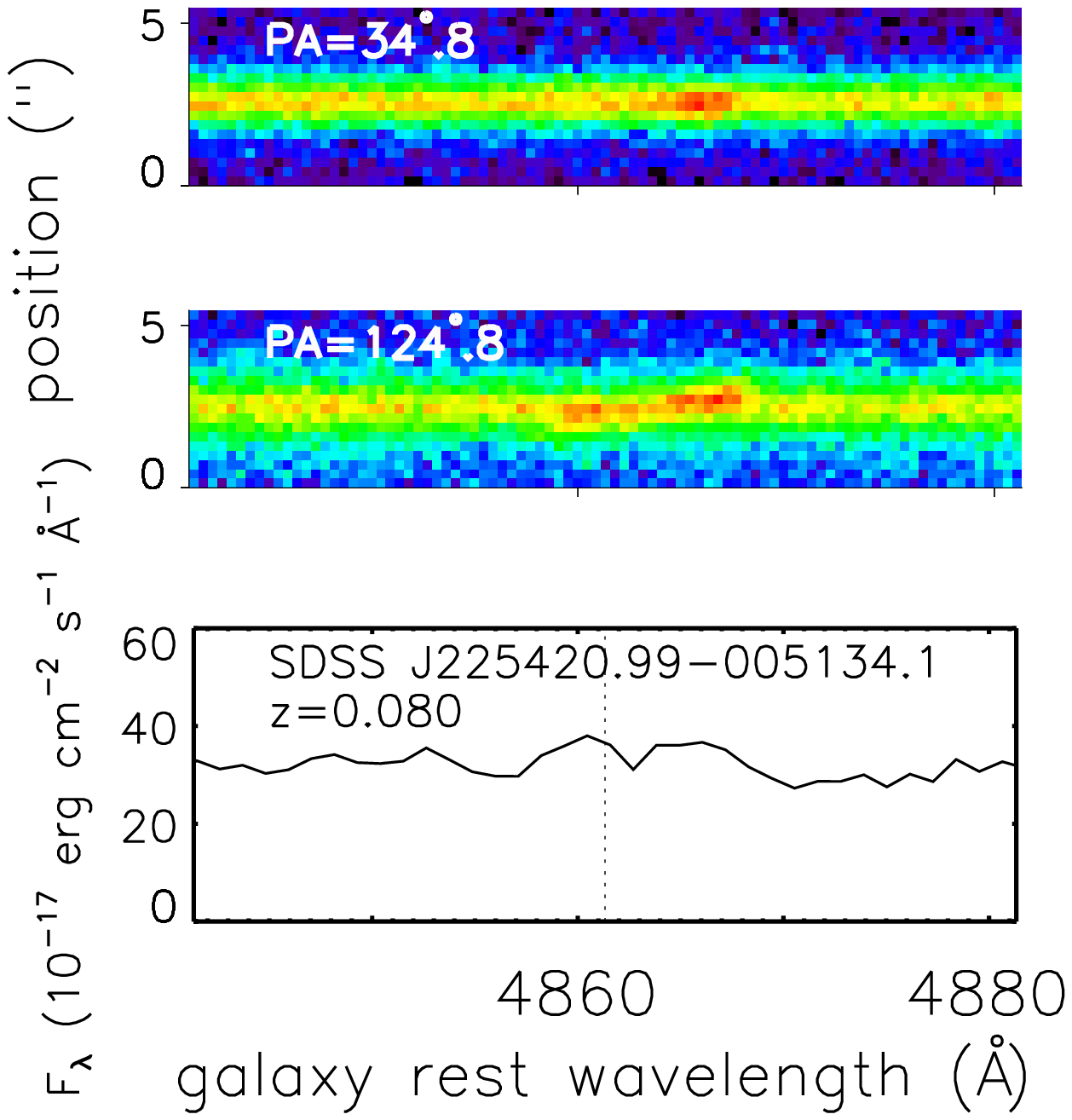}}
\hspace{-0.7cm}
\subfigure{ \includegraphics[height=3.44cm]{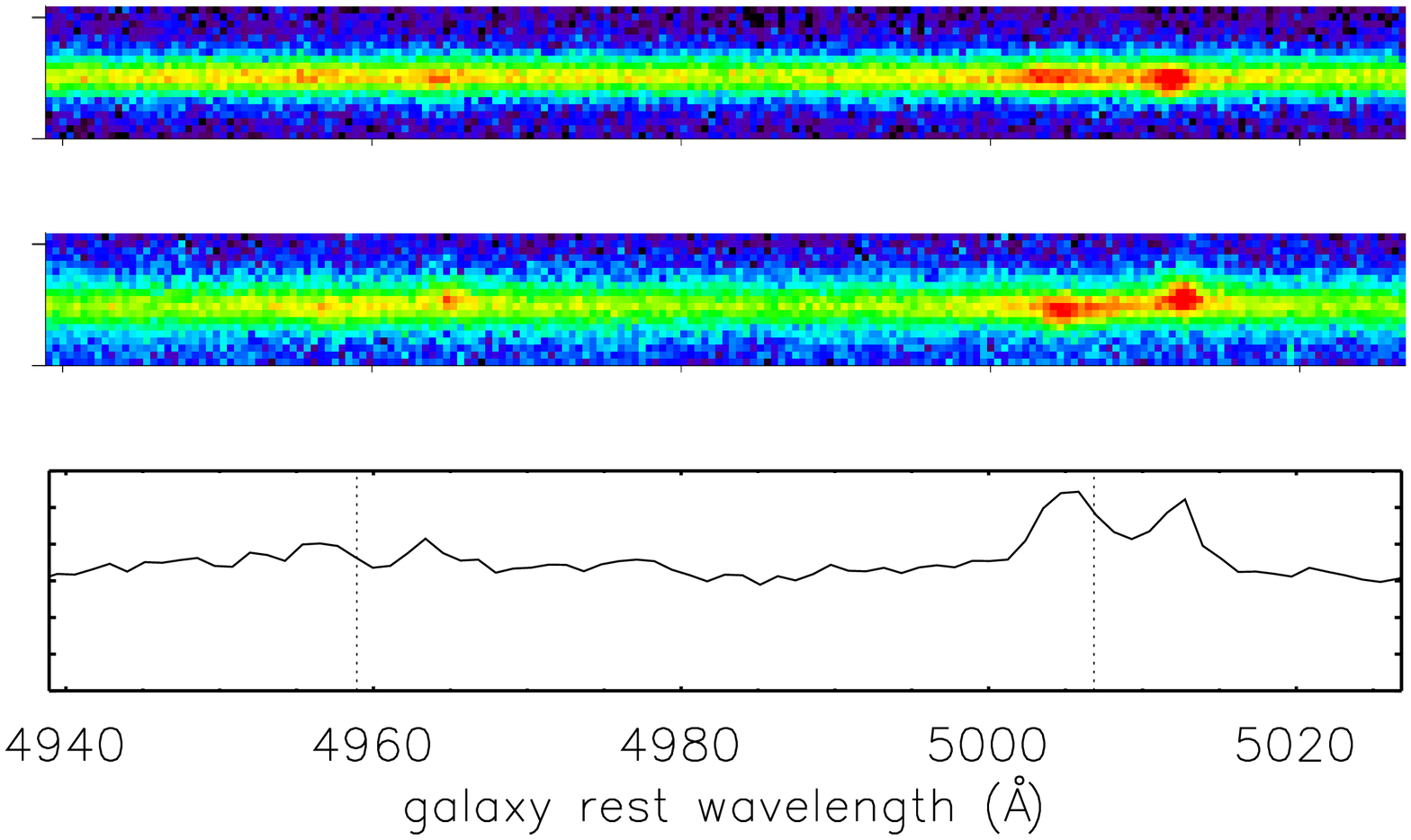}} 
\subfigure{ \includegraphics[height=3.35cm]{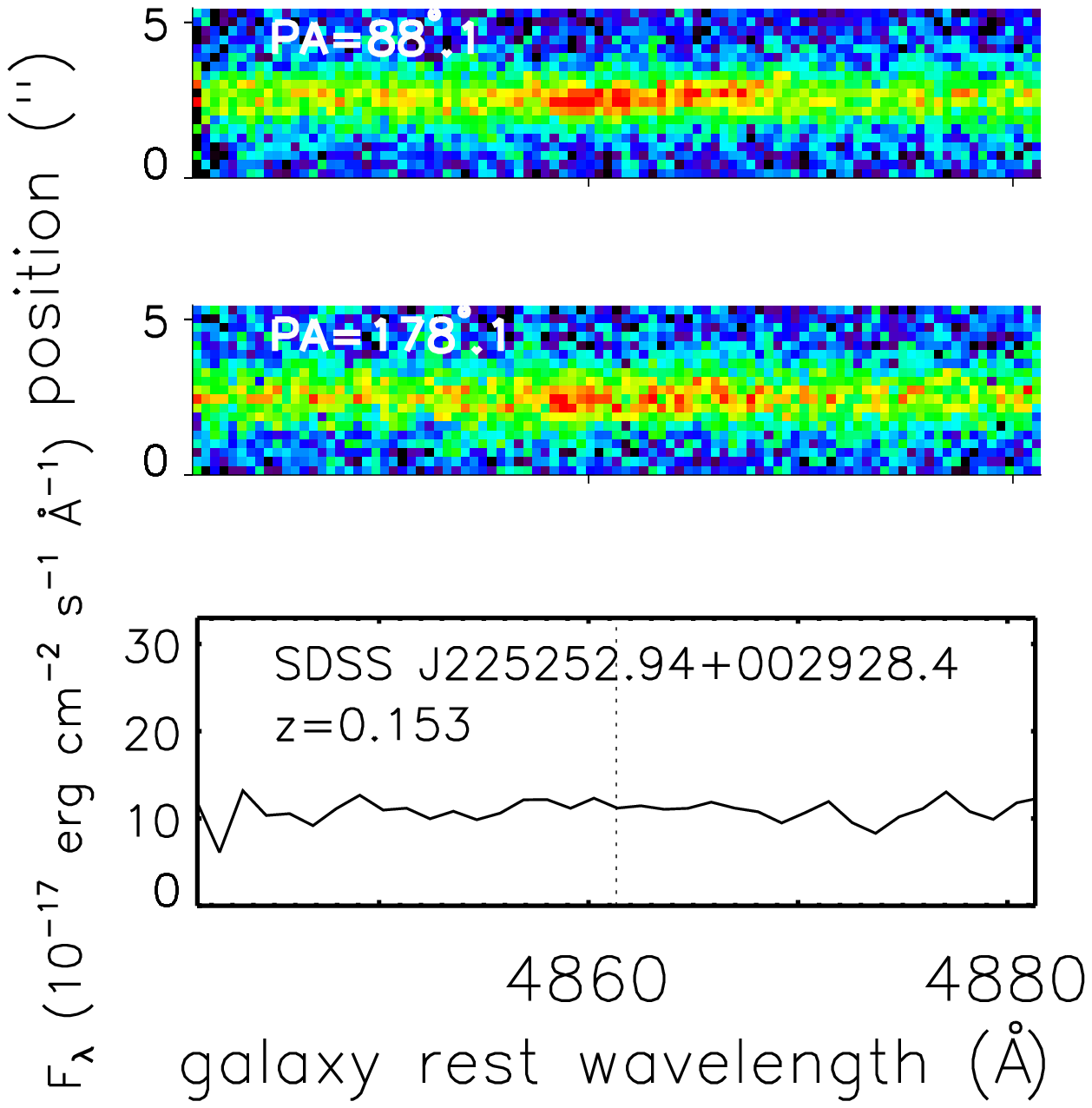}}
\hspace{-0.7cm}
\subfigure{ \includegraphics[height=3.35cm]{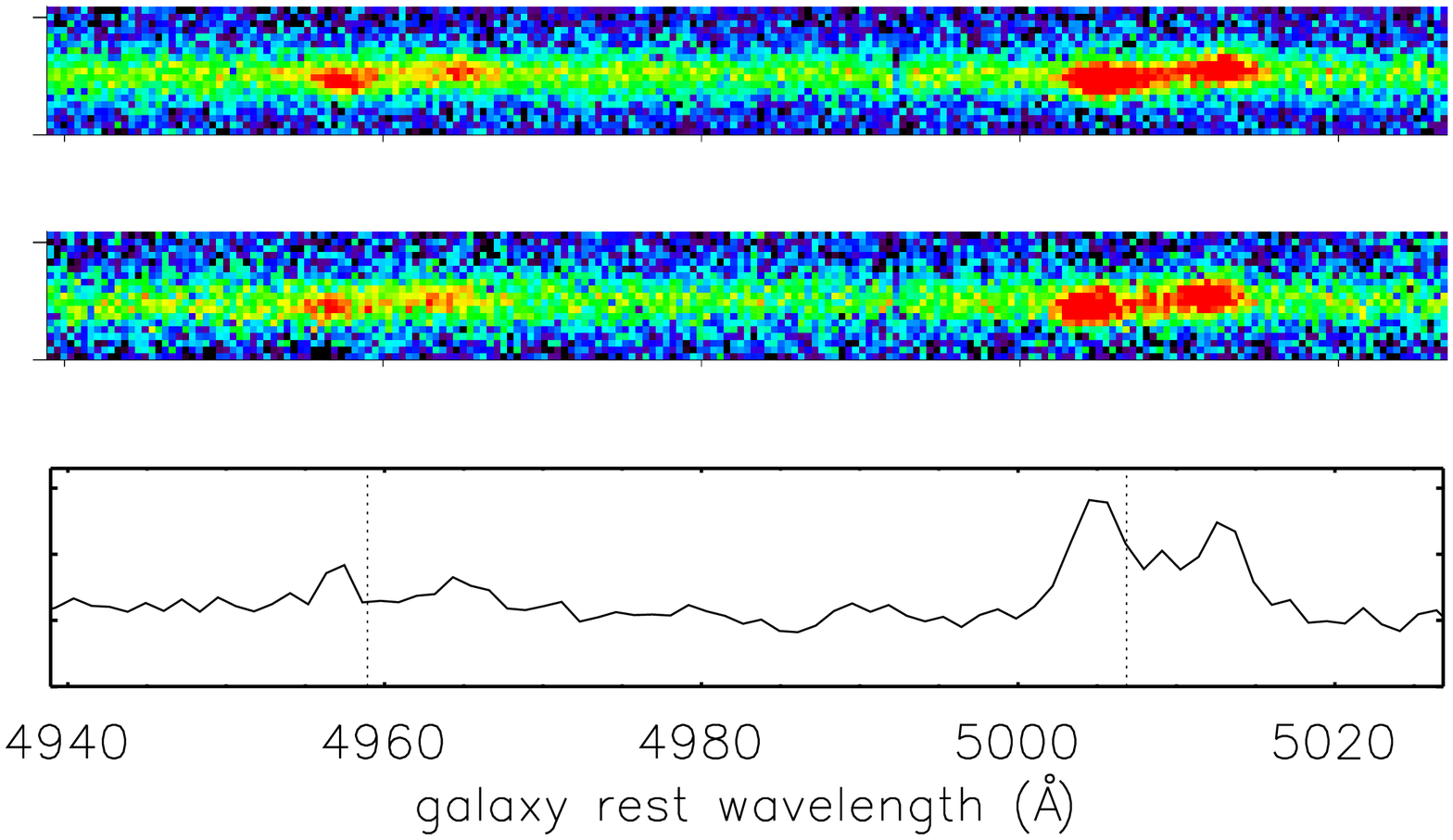}}
\caption{\footnotesize{Segments of the two-dimensional long-slit spectra and one-dimensional SDSS spectra of five example double-peaked AGNs that exhibit two spatially compact emission components.  Each object was observed at two roughly orthogonal position angles, as labeled in the two-dimensional spectra.  The spectra are shifted to the rest frame of the host galaxy, and the dotted vertical lines denote the rest wavelengths of \hbn, \oiiiwwn, and \oiiiwn. Night-sky emission features have been subtracted from the two-dimensional spectra.}}
\label{fig:compact}
\end{figure}

Using the $r$-band Stokes parameters $Q_r$ and $U_r$ from the SDSS photometric pipeline \citep{ST02.3}, we compute the ellipticity $e=\sqrt{Q_r^2+U_r^2}$ for each object in our sample.  While adaptive moments can be used for more refined measurements of galaxy shapes and are used in weak lensing studies \citep{BE02.1}, for our purpose of measuring the overall ellipticity of a galaxy's light the Stokes parameters are the most appropriate.

\section{long-slit Spectral Analysis}

\subsection{Spatial Separation Measurements}
\label{measure}

For each two-dimensional long-slit spectrum, we determine the projected spatial separation between the two \oiiiw emission features by measuring the spatial centroid of each emission component individually.  We measure the spatial centroid of an emission component by first centering a 2 \AA $\,$ (rest-frame) wide window on the emission.  Next, at each spatial position we sum the flux, weighted by the inverse variance, over all wavelengths within the window.  We place a 10 pixel window around the emission, and define the window's center to be the spatial position where the summed flux is maximum.  We then fit a quadratic to the summed flux to locate a peak, define a narrow window centered on the peak flux, and compute the line centroid within this window. We derive the error on this spatial centroid by repeatedly adding noise to the spectrum drawn from a Gaussian with the variance of the pixels in the window and redoing all centroid measurements. 

For a given target, we measure the projected spatial separation between the two \oiiiw emission components at both position angles observed, and then combine the two spatial separations to yield the spatial offset between the two emission components on the sky and the position angle of their offset.  If observations at position angles $\theta_1$ and $\theta_2$ yield spatial separation measurements of $x_1$ and $x_2$, respectively, then the true position angle $\theta_{\rm{sky}}$ of the two emission components on the sky is determined through the numerical solution of

\begin{equation}
x_1 \cos(\theta_{\mathrm{sky}}-\theta_2) = x_2 \cos (\theta_{\mathrm{sky}} - \theta_1) \, .
\end{equation}

With $\theta_{\mathrm{sky}}$ determined above, the spatial separation of the two emission components on the sky is then $\Delta x=x_1/\cos(\theta_{\mathrm{sky}}-\theta_1)$ or equivalently, $\Delta x=x_2/\cos(\theta_{\mathrm{sky}}-\theta_2)$.

\subsection{Spatial Extent of the Emission}
\label{extent}

We also classify by eye whether the AGN emission components in the long-slit spectra are spatially ``compact" or ``extended".  We find that each emission feature displays two components, and if both components appear spatially distinct and compact at both position angles observed we label the object ``compact".  If one or both of the components appear spatially extended or diffuse at one or both position angles, we label the object ``extended".  This type of extended structure is known to be produced by gas kinematics, outflows, or jets (e.g., \citealt{RO10.1,FI11.1}), though we note it could also be produced by dual AGNs where one or both of the AGNs also has visible emission from gas kinematics, outflows, or jets. 

\begin{figure}
\centering
\subfigure{\includegraphics[height=5.1cm]{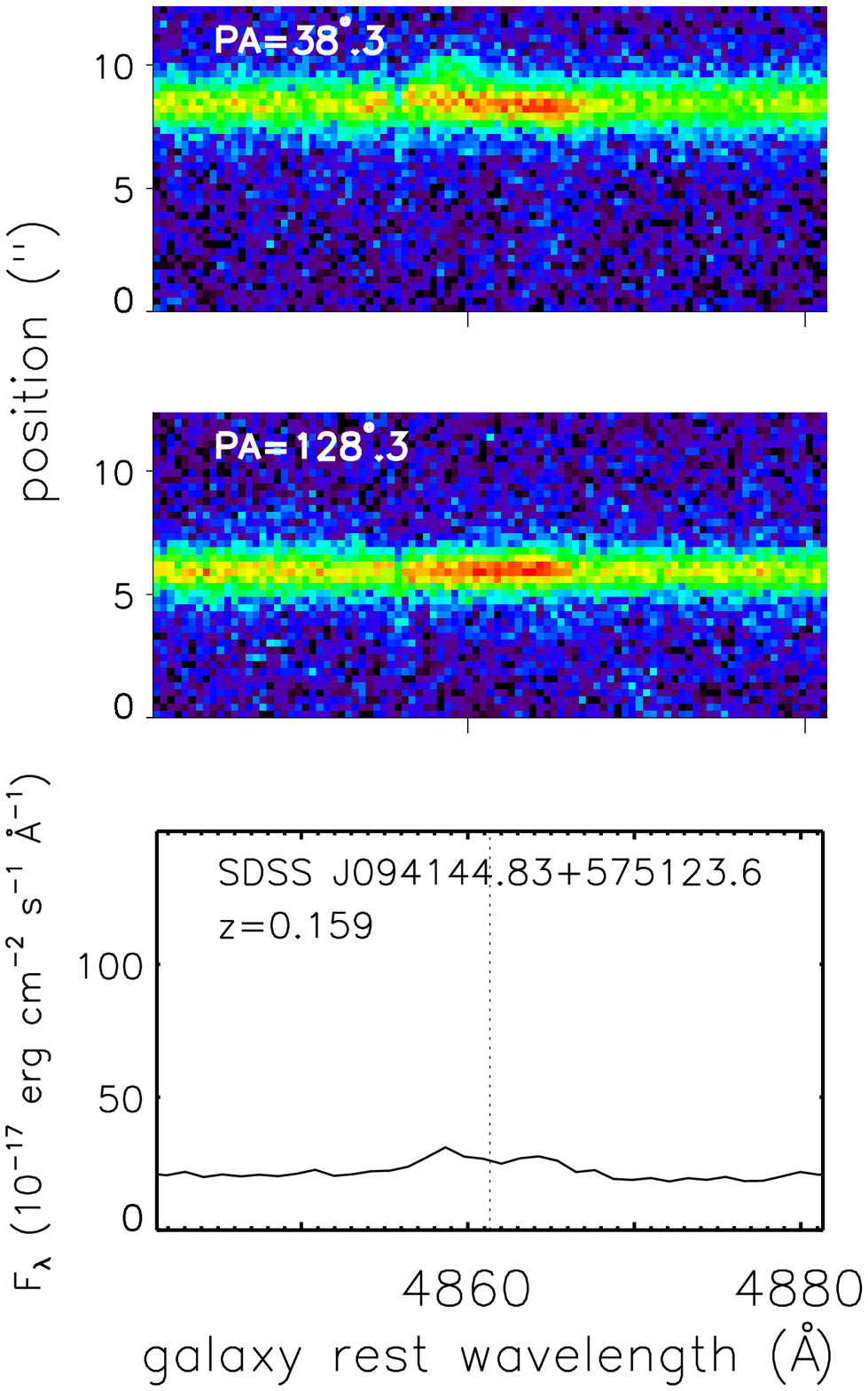}}
\hspace{-0.7cm}
\subfigure{\includegraphics[height=5.1cm]{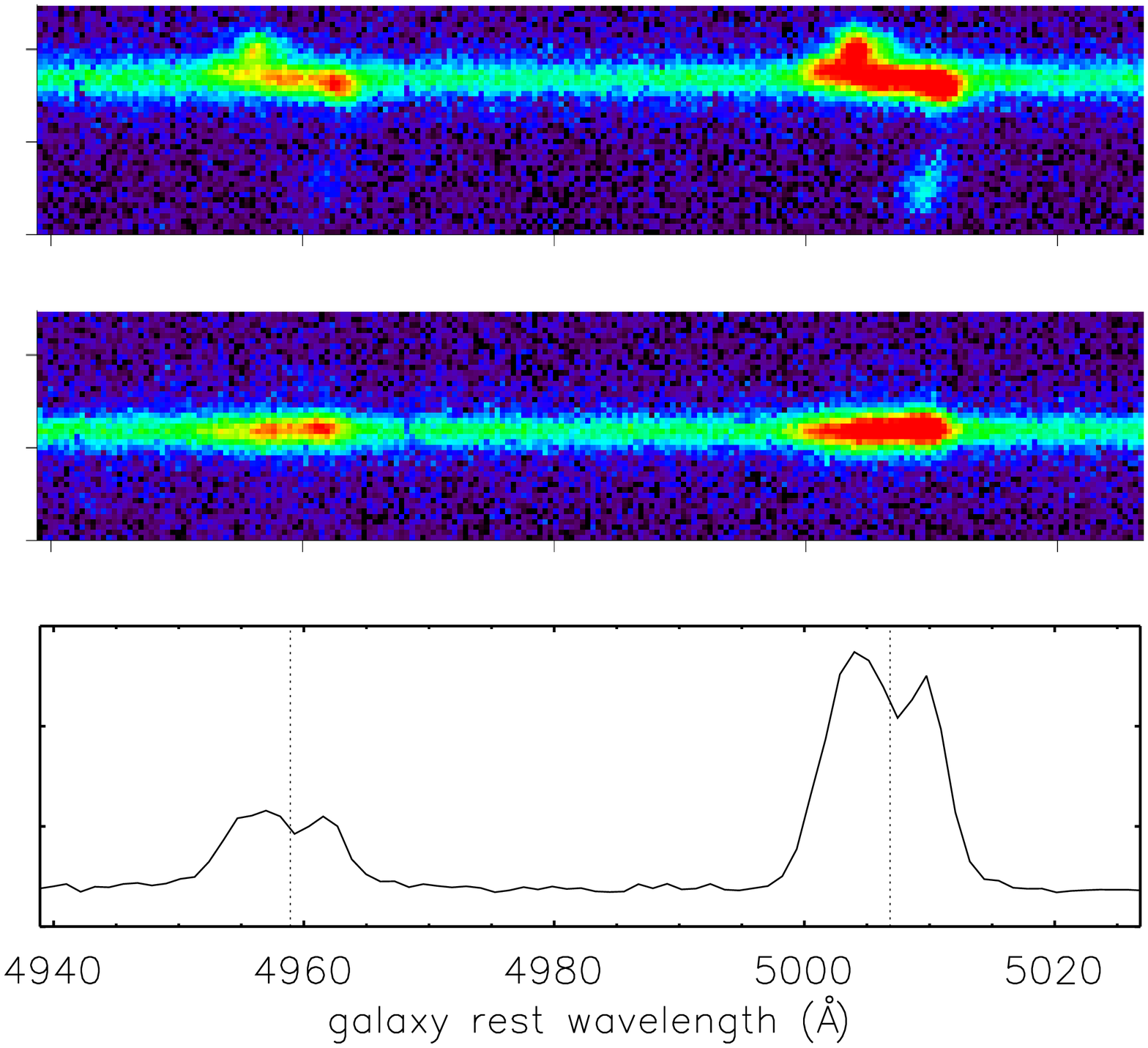}} 
\subfigure{\includegraphics[height=4.cm]{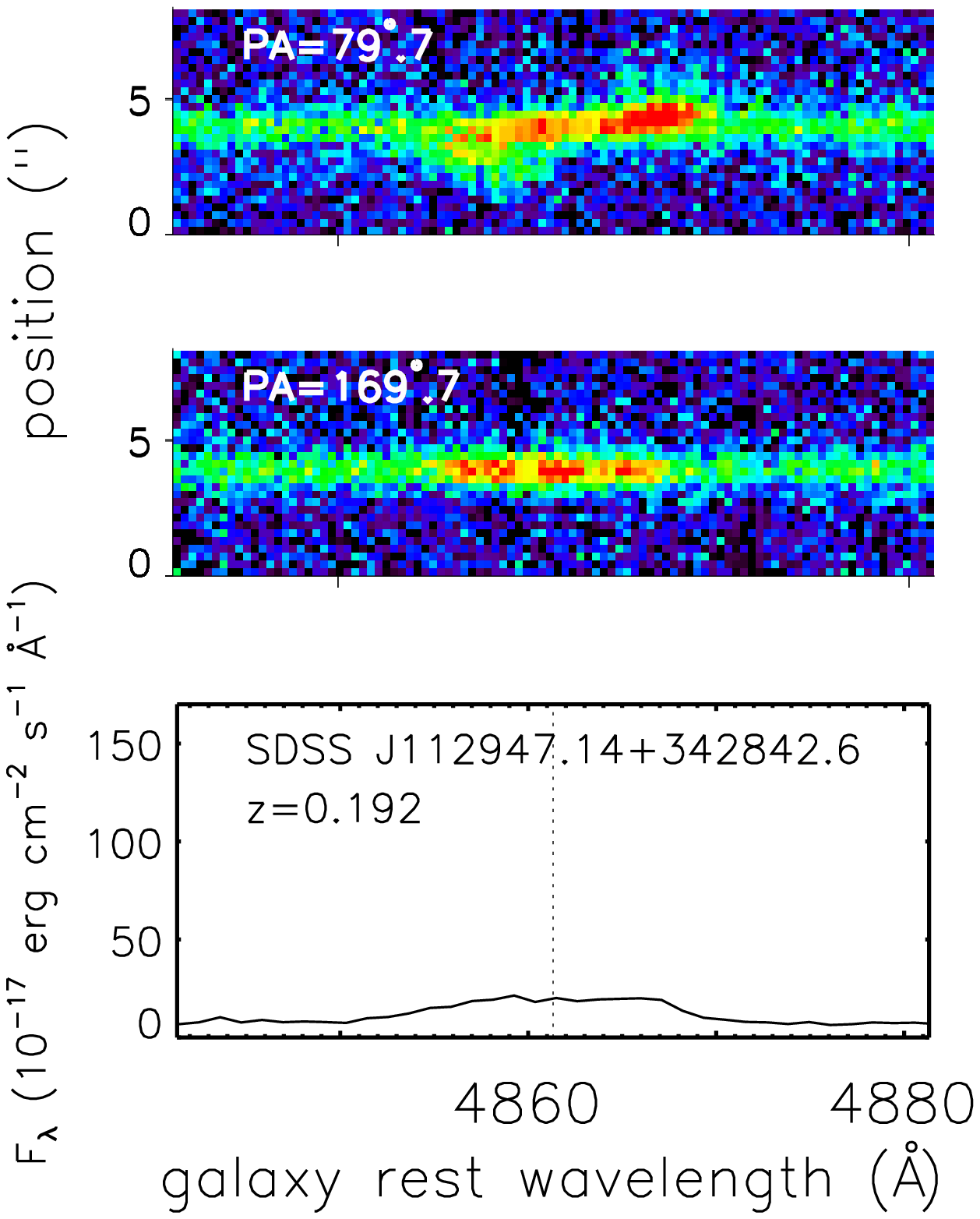}}
\hspace{-0.7cm}
\subfigure{\includegraphics[height=4.cm]{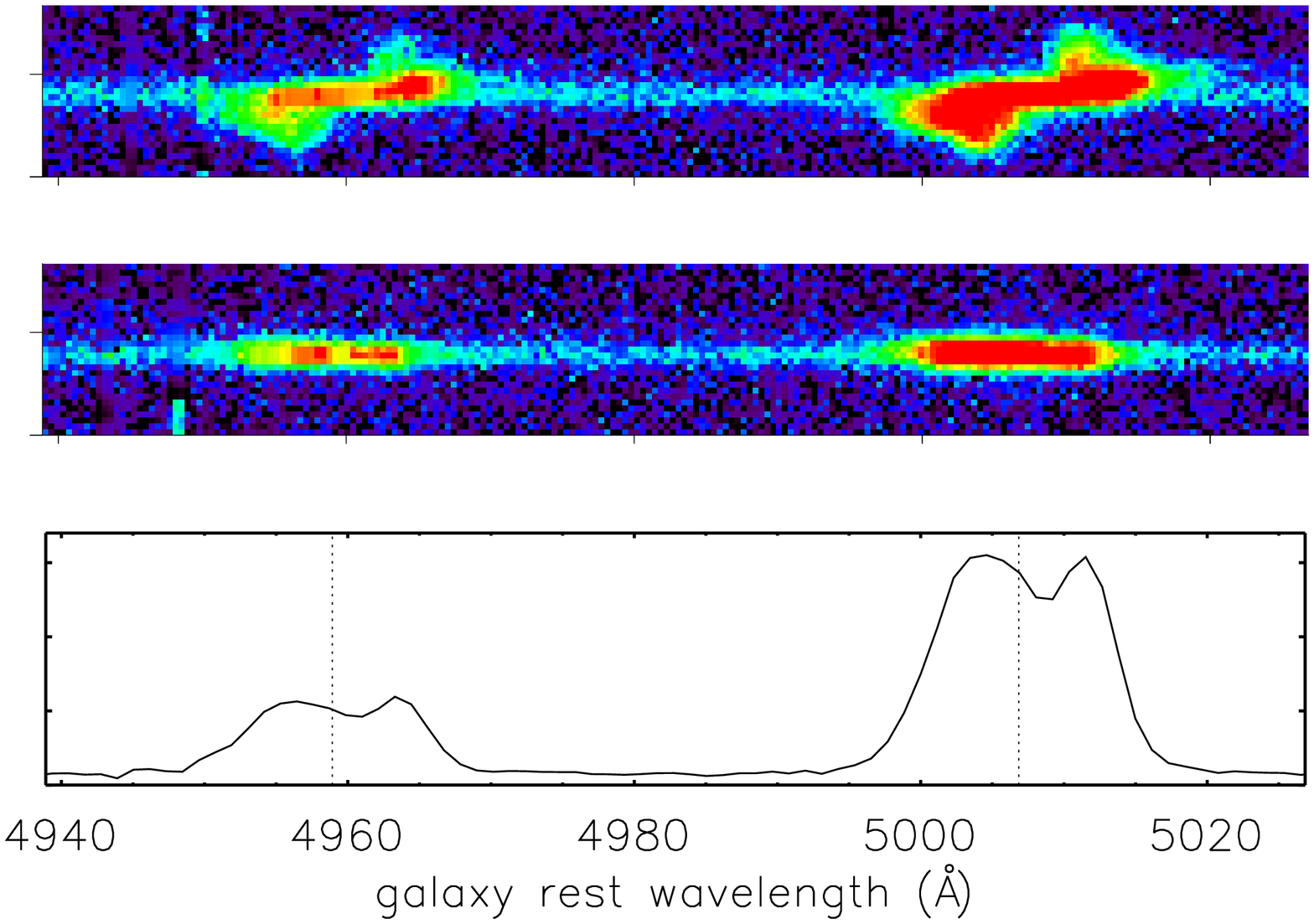}} 
\subfigure{ \includegraphics[height=8.6cm]{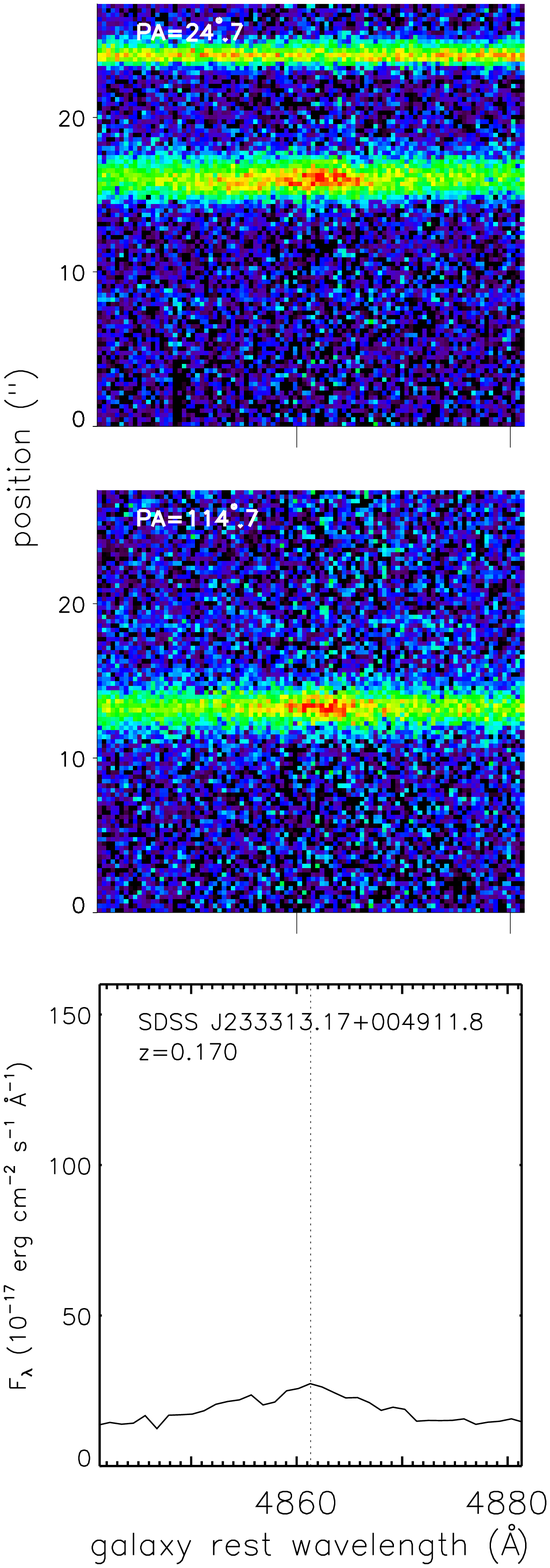}}
\hspace{-0.7cm}
\subfigure{ \includegraphics[height=8.6cm]{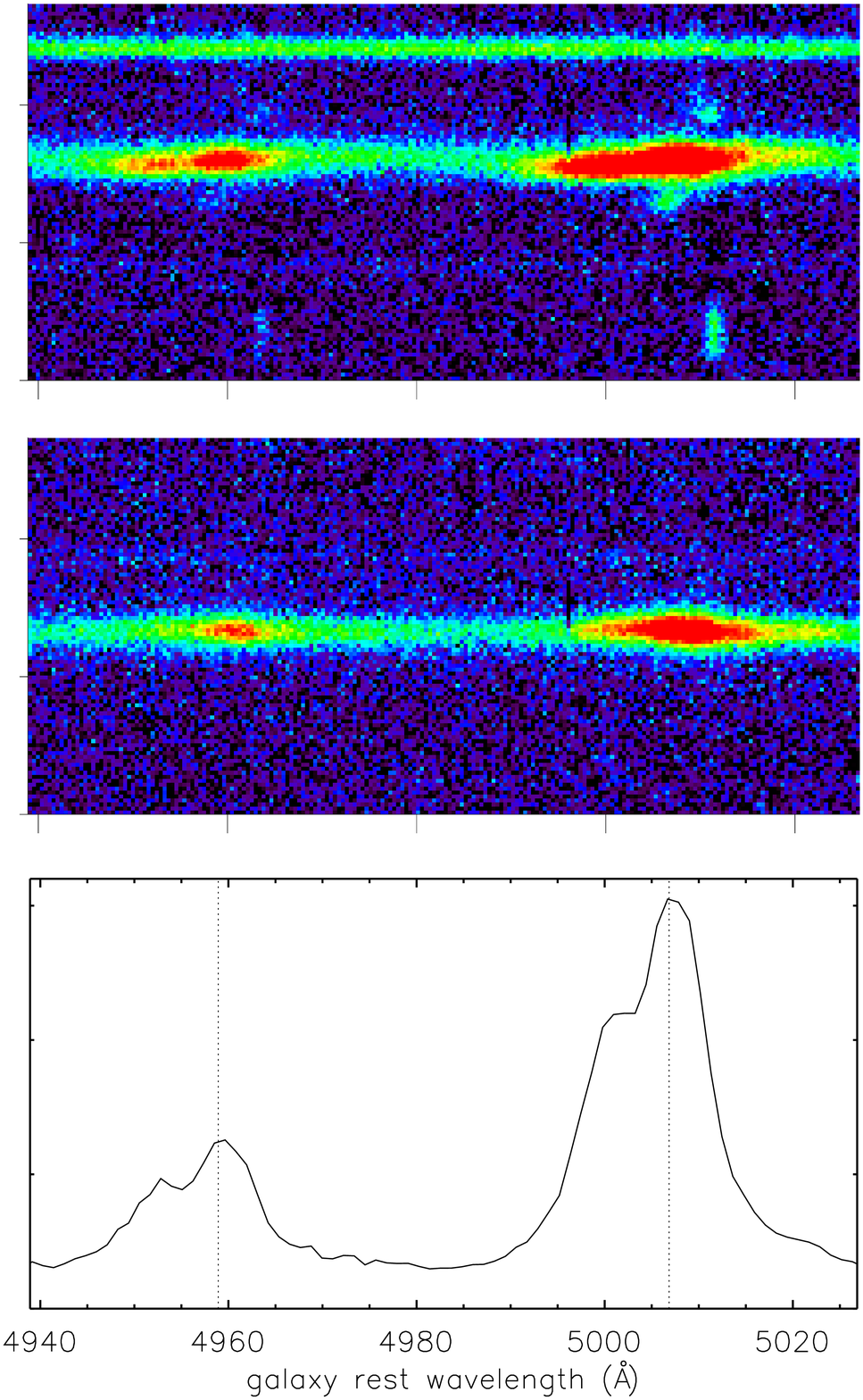}}
\caption{Same as Figure~\ref{fig:compact}, but for three example double-peaked AGNs that exhibit spatially extended emission.}
\label{fig:extended}
\end{figure}

We find 47 spatially compact objects and 34 spatially extended objects, corresponding to $58^{+5}_{-6} \, \%$ and $42^{+6}_{-5} \,\%$ of the sample, respectively.  Figure~\ref{fig:compact} shows examples of the spatially compact object spectra, while Figure~\ref{fig:extended} shows examples of the spatially extended object spectra.

Sections \ref{multiwavelength} and \ref{unusual} address other observations of our sample and some objects that we find have unusual long-slit spectra; for results for the complete sample skip to Section \ref{results}.

\section{Multiwavelength Observations from the Literature}
\label{multiwavelength}

In an effort to classify the nature of the double-peaked \oiiiw emission lines in SDSS spectra, some authors have conducted follow-up observations of subsamples of these objects in optical, near-infrared (NIR), radio, and X-ray.  These observations aimed to test the dual AGN hypothesis by resolving systems with two AGNs or two stellar components, where the stellar components are presumed to be remnants of the progenitor galaxies from the merger that produced the dual AGNs.
Here we summarize these observations and their overlap with our sample (see also Table~\ref{tbl-2}).

The follow-up optical slit spectroscopy and NIR imaging of 31 Type 2 Seyferts with double-peaked \oiiiw \citep{LI10.2, SH11.1} includes 14 objects in our sample.  None of our objects exhibit two stellar components in the NIR imaging.  However, in three of our objects we measure that the emission components in the long-slit spectra are separated by less than the $\sim0\farcs4$ spatial resolution of the NIR imaging, such that stellar nuclei coincident with the emission components would not be spatially resolved in the NIR imaging.  In the remaining 11 objects, stellar nuclei that are spatially coincident with the double emission components could be resolved in the NIR imaging, so either they have no secondary stellar component at these spatial separations, or the secondary stellar nucleus is obscured or too faint to be visible in the NIR imaging.

Follow-up adaptive optics (AO) imaging strived to resolve smaller-separation stellar components in 55 galaxies with double-peaked \oiiiw \citep{FU11.1,RO11.1,FU12.1}.  In addition to obtaining integral field unit (IFU) spectroscopy of 42 galaxies with double-peaked \oiiiwn, \cite{FU12.1} also analyzed high-resolution Advanced Camera for Surveys/WFC2 archival {\it Hubble Space Telescope} ({\it HST}) images of six additional double-peaked galaxies.  These IFU, AO, and {\it HST} observations include 34 objects in our sample, of which seven show double stellar components in AO.  

\begin{figure}[!t]
\centering
\subfigure{\includegraphics[height=4.1cm]{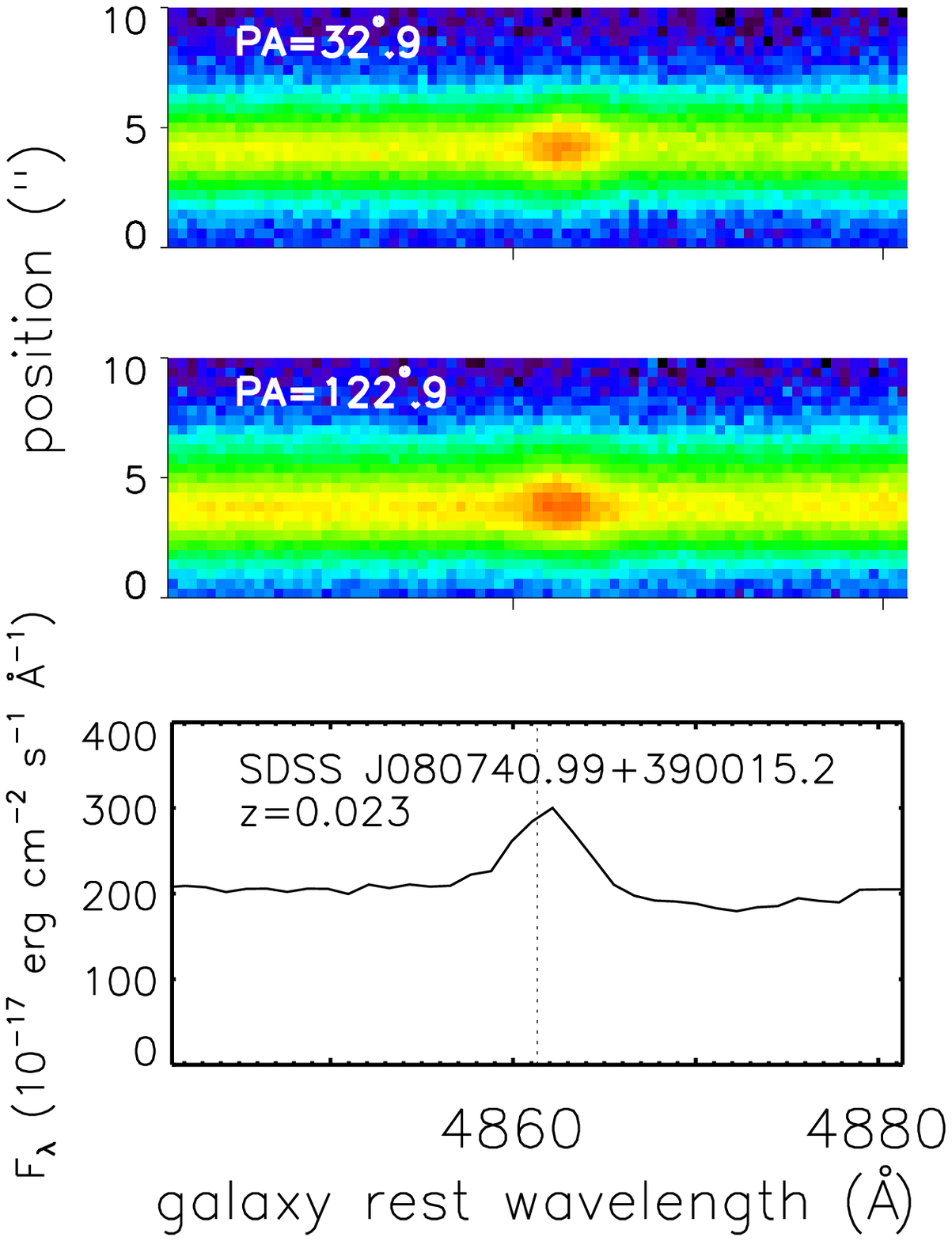}}
\hspace{-0.7cm}
\subfigure{\includegraphics[height=4.1cm]{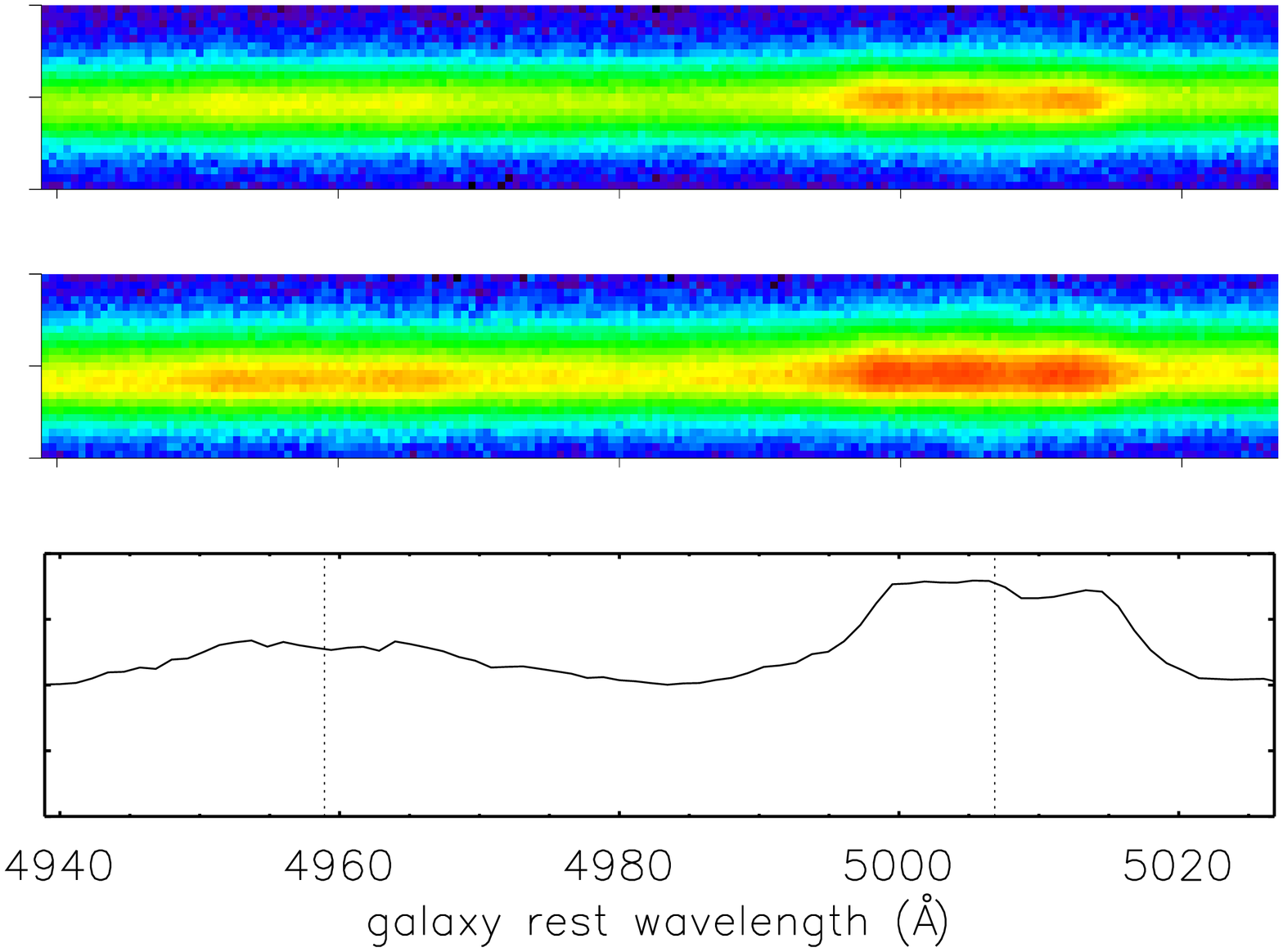}} 
\subfigure{\includegraphics[height=10.4cm]{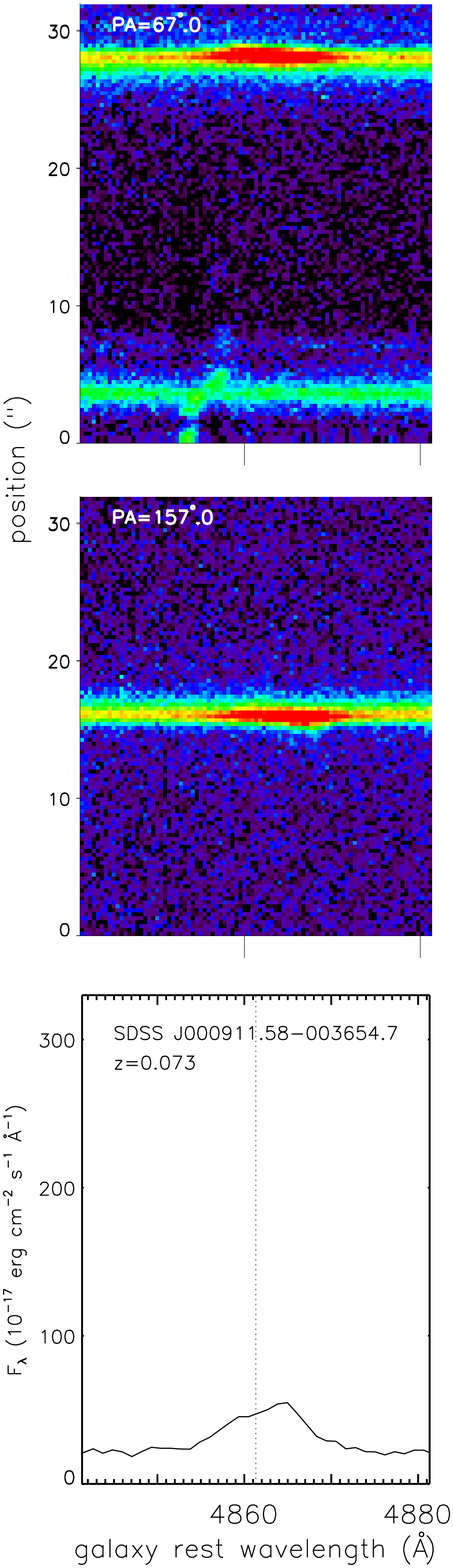}}
\hspace{-0.7cm}
\subfigure{\includegraphics[height=10.4cm]{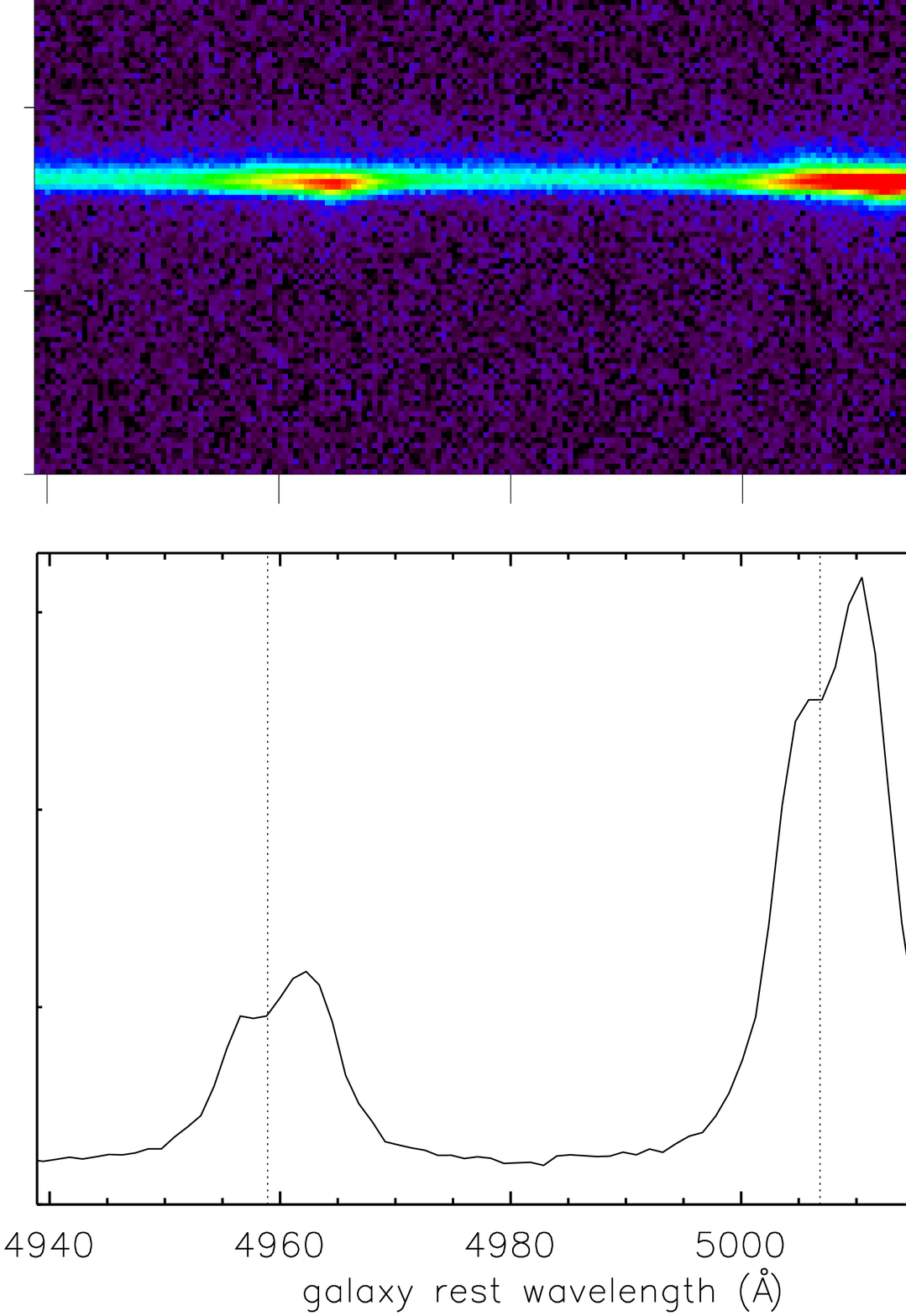}} 
\subfigure{\includegraphics[height=5.5cm]{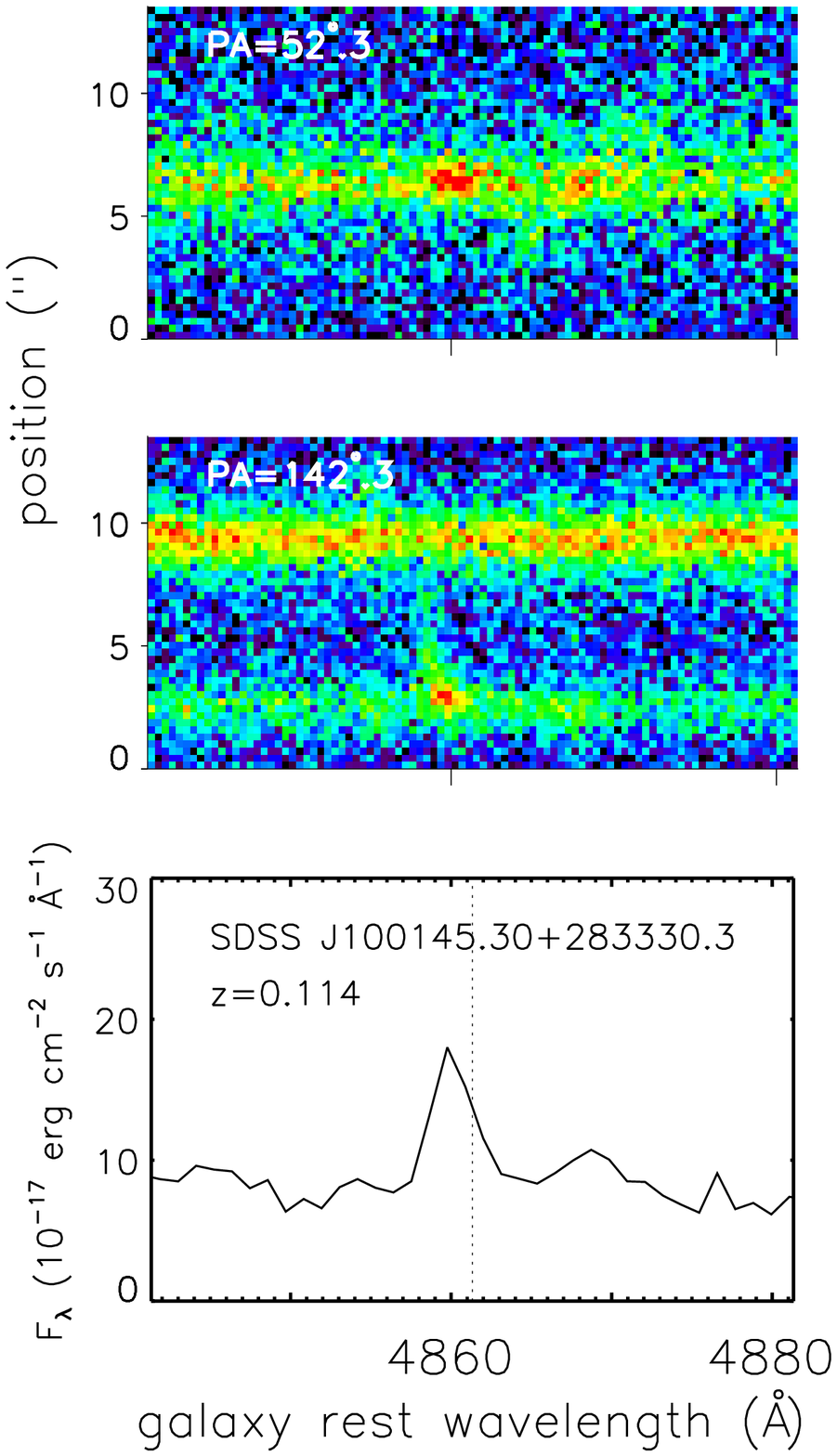}}
\hspace{-0.7cm}
\subfigure{\includegraphics[height=5.5cm]{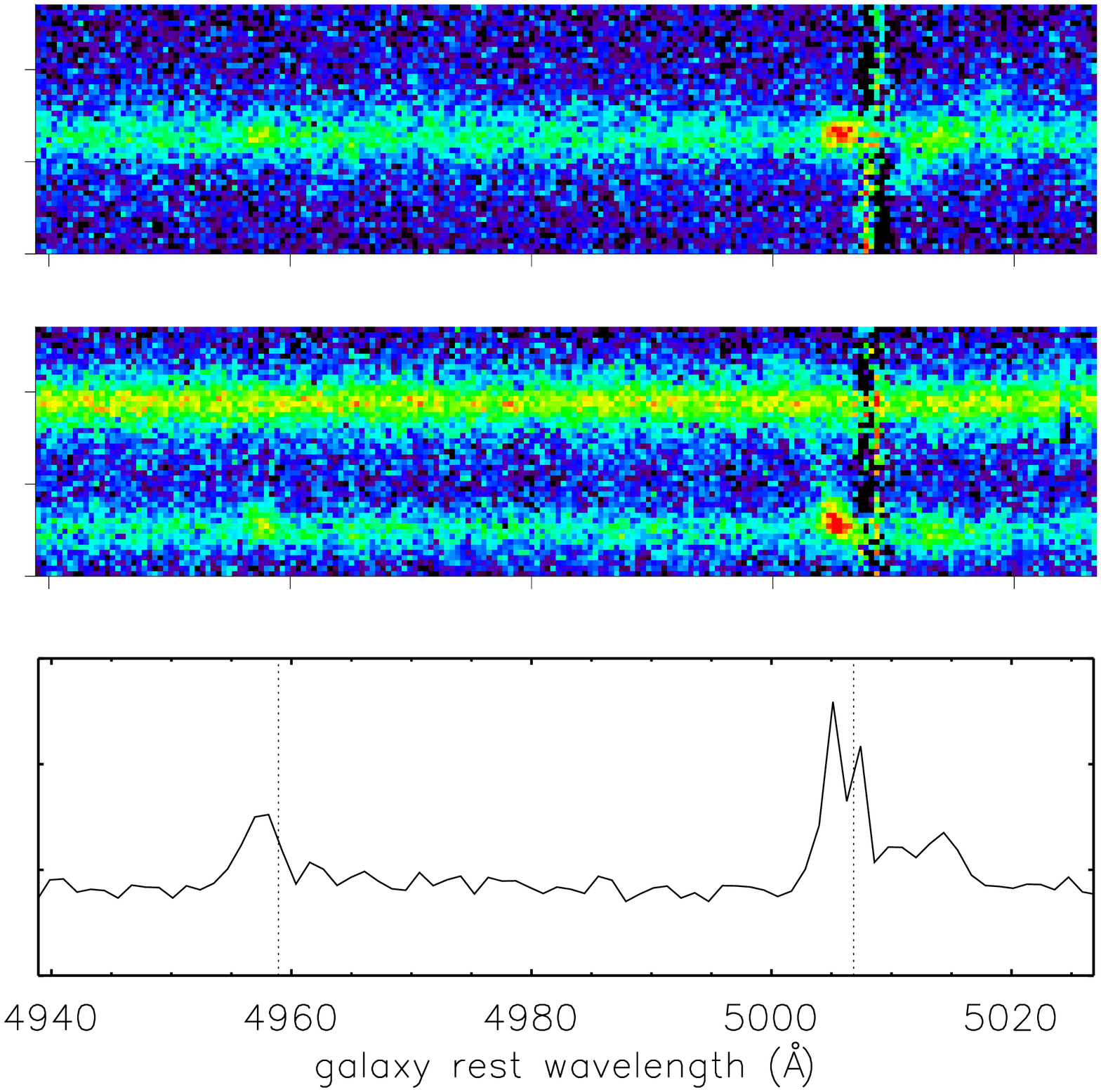}}
\caption{\footnotesize{Same as Figure~\ref{fig:compact}, but for three noteworthy galaxies in our sample that have unusual emission features.  SDSS J080740.99+390015.2 (top) has triple-peaked narrow lines and single-peaked broad lines; SDSS J000911.58-003654.7 (middle) has a companion spiral galaxy that may be at a similar redshift; and SDSS J100145.30+283330.3 (bottom) has spatially extended emission linking it to its companion galaxy, which is a possible signature of gas outflow or inflow.}}
\label{fig:unusual_spectra}
\end{figure}

Although position angles between the stellar components are not reported in the literature \citep{FU11.1,RO11.1,FU12.1}, we estimate the position angles for these seven objects based on our best estimates of the bright centroid of each component in the published images.  In Section~\ref{ao}, we compare these position angle estimates, as well as the reported spatial separations, to the position angles and spatial separations measured from our long-slit spectroscopy.  

Further, Very Long Baseline Array (VLBA) observations of 11 Type 2 Seyferts that have double-peaked \oiiiw as well as Faint Images of the Radio Sky at Twenty-cm (FIRST) detections reveal compact radio emission in only two objects \citep{TI11.1}.  One (SDSS J151709.21+335324.7) of these two galaxies is part of our sample, and this galaxy also has a radio jet that suggests a radio jet-driven outflow produces the double-peaked \oiiiw lines \citep{RO10.1}. None of the VLBA observations show double radio cores, although Expanded Very Large Array observations of another double-peaked AGN (not in our sample) revealed two radio cores that confirmed it as a dual AGN system \citep{FU11.3}.  We also find no significant correlation between FIRST detections and other properties of double-peaked AGNs that are measured here, such as line-of-sight velocity separation and projected physical separation.

Finally, {\it Chandra} observations of the double-peaked AGN in SDSS J171544.05+600835.7 show two X-ray sources with separation $\Delta x=1.9$ $h_{70}^{-1}$ kpc and orientation $\theta_{\rm sky}=147^\circ$, which are consistent with the separation and orientation measured between the double \oiiiw sources in the long-slit spectra \citep{CO11.2}.  While this is strong evidence for dual AGNs, additional follow-up observations are required for confirmation of dual AGNs in this system.

\begin{figure}[!t]
\centering
\subfigure{\includegraphics[height=5.2cm]{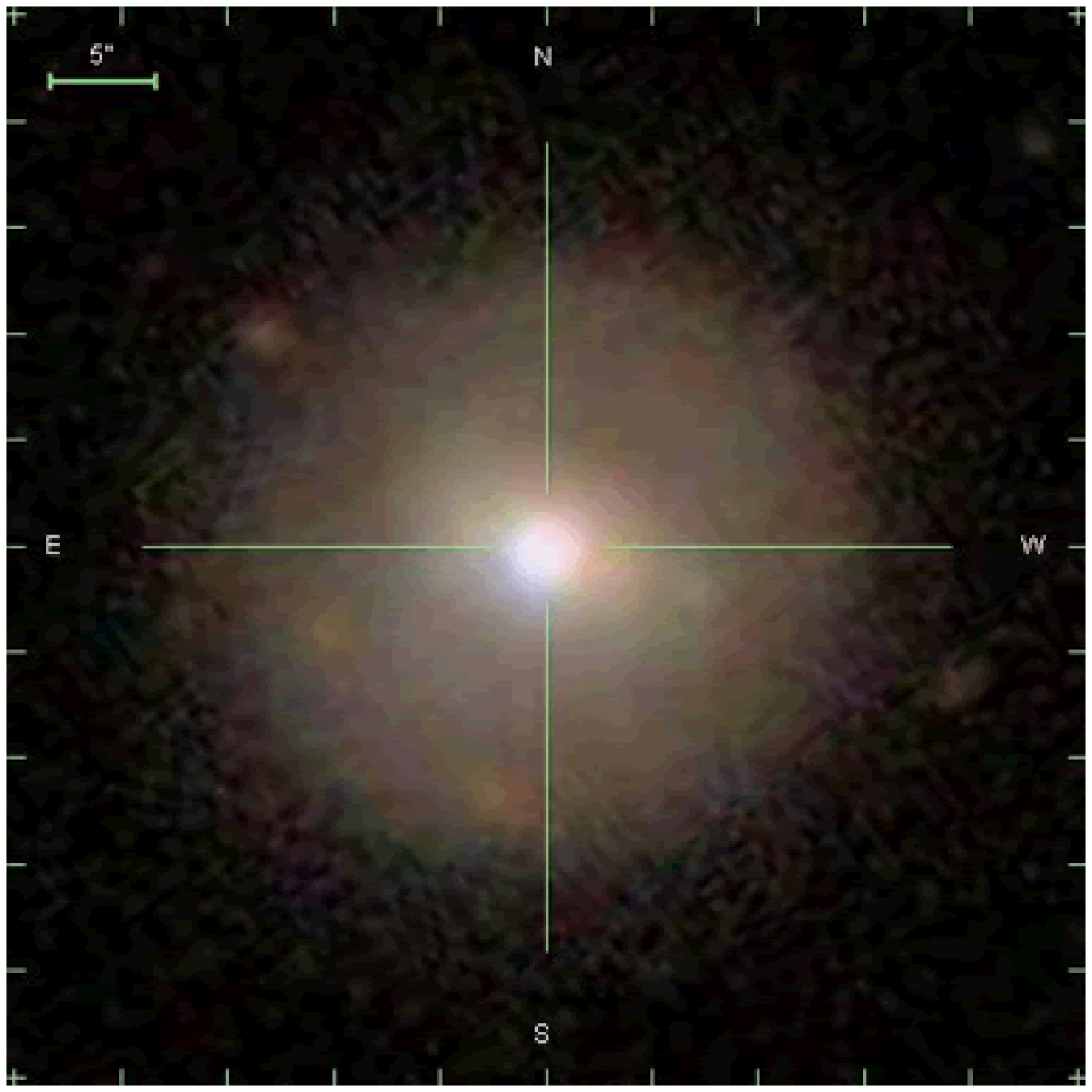}} 
\subfigure{\includegraphics[height=5.2cm]{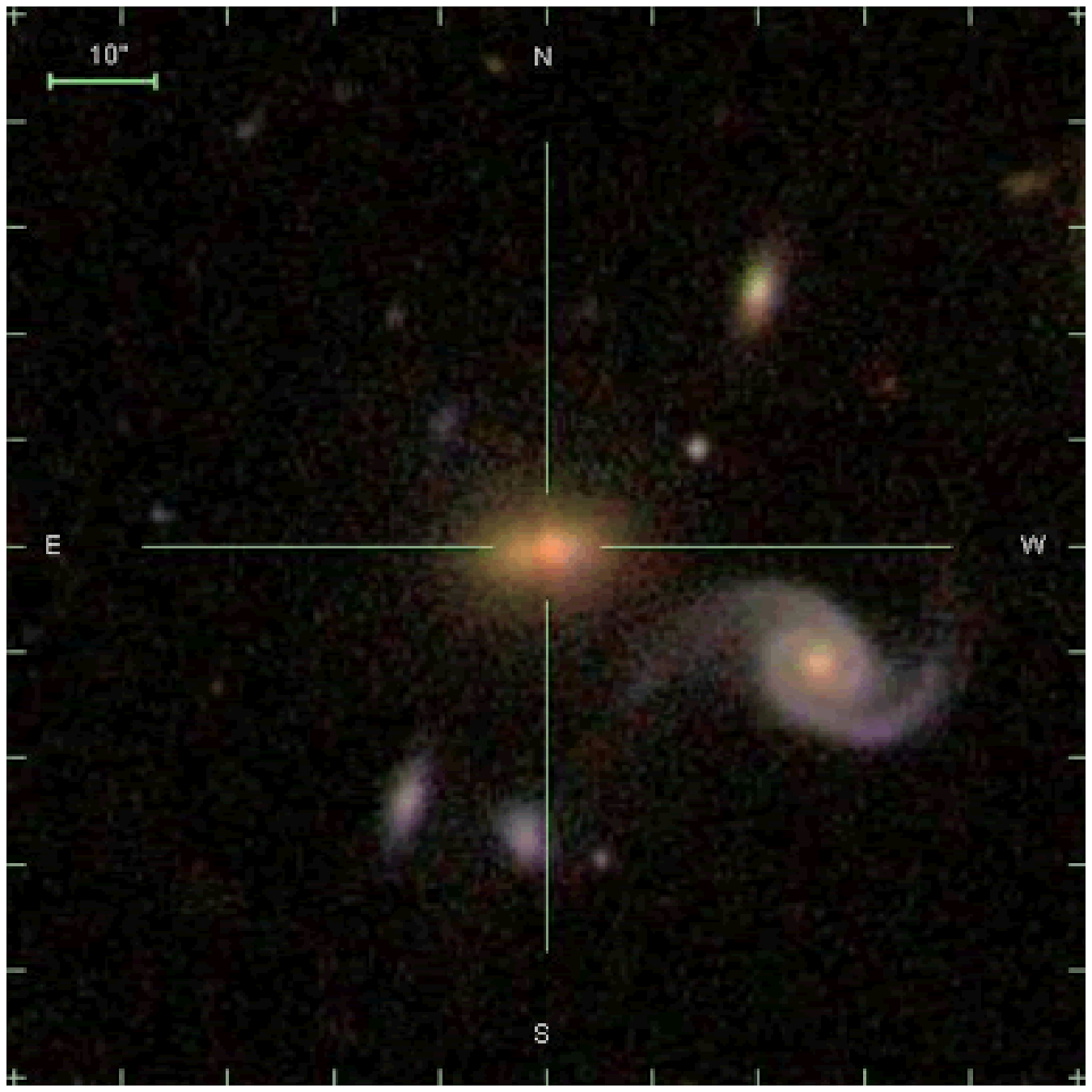}} 
\subfigure{\includegraphics[height=5.2cm]{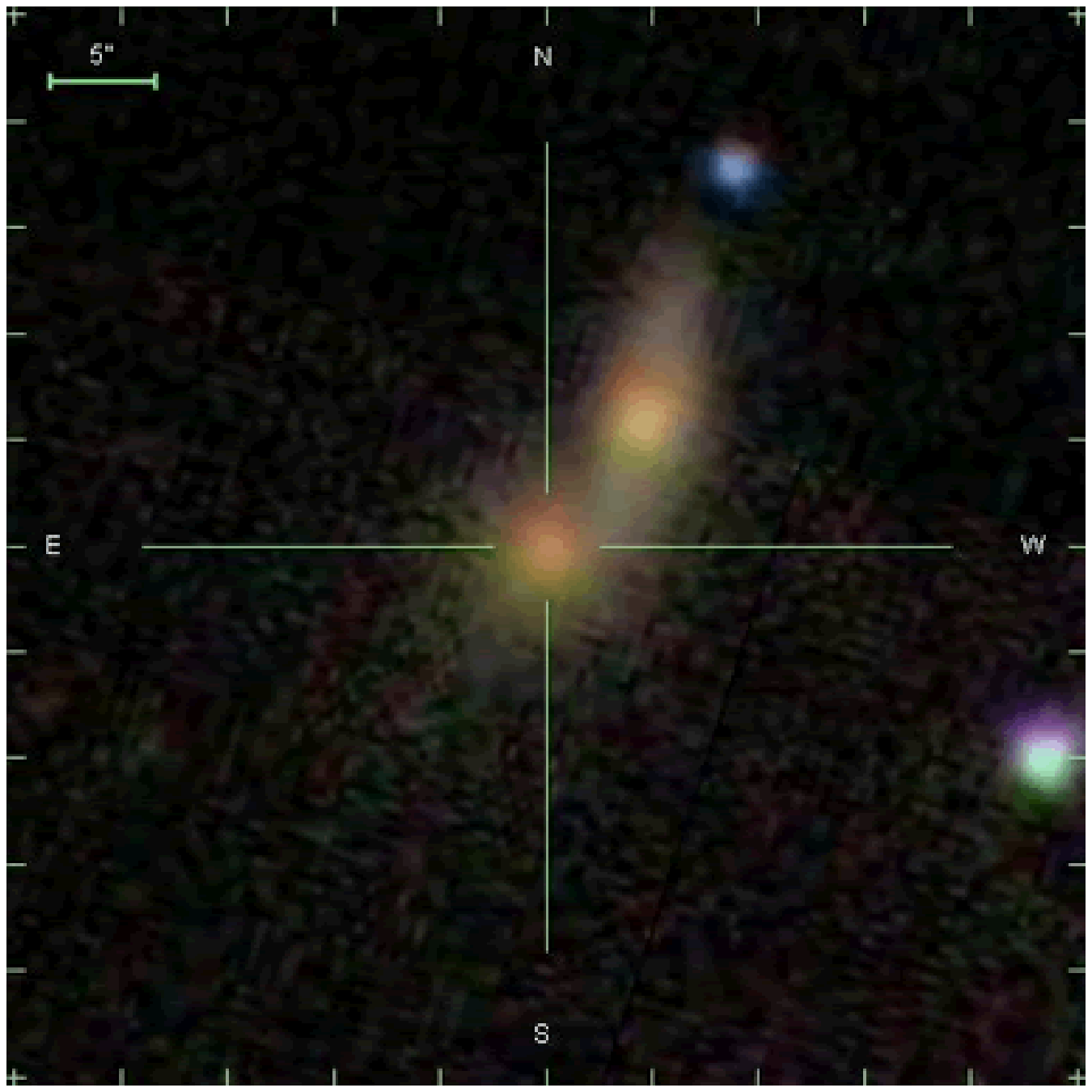}} 
\caption{SDSS images of the three galaxies whose spectra are shown in Figure~\ref{fig:unusual_spectra}.  SDSS J080740.99+390015.2 (top) may host AGN outflows or jets; SDSS J000911.58-003654.7 (middle) has a companion spiral galaxy visible to the southwest; and SDSS J100145.30+283330.3 (bottom) has a companion galaxy visible to the northwest.}
\label{fig:unusual_images}
\end{figure}

\section{Unusual Spectra}
\label{unusual}

Three of our targets have particularly unusual and noteworthy spectra, which we discuss briefly here.  Figure~\ref{fig:unusual_spectra} shows the spectra and Figure~\ref{fig:unusual_images} shows the corresponding SDSS images of these galaxies. 

\subsection*{SDSS J080740.99+390015.2} 

Although this galaxy was classified as a double-peaked AGN in \cite{WA09.1}, its spectra display triple-peaked narrow lines and single-peaked broad lines (Figure~\ref{fig:unusual_spectra}, top), a line profile that could be explained by outflows or jets that affect the NLR only.  We exclude this galaxy from the analysis in this paper, as we desire a homogeneous sample of double-peaked AGNs.  

\begin{figure}[!t]
\begin{center}
\includegraphics[height=3.5in]{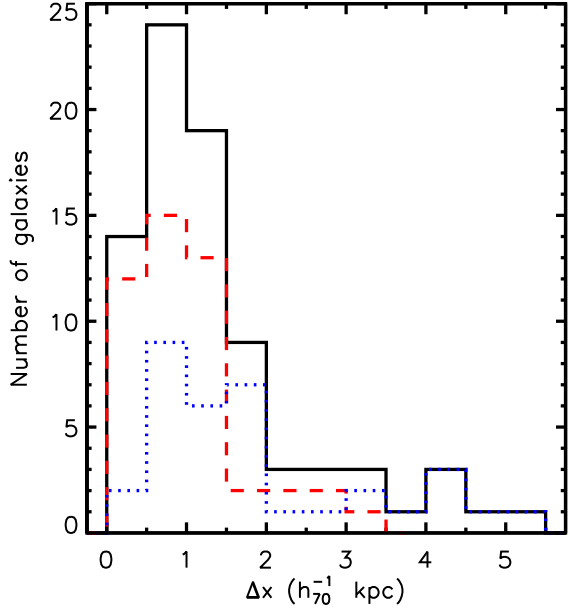}
\end{center}
\caption{Histograms of projected physical separations on the sky measured between double \oiiiw emission components in the long-slit spectra for the entire sample (black solid), objects with spatially compact emission components (red dashed), and objects with spatially extended emission components (blue dotted).  The median physical separation is 1.1 $h^{-1}_{70}$ kpc.  See Section~\ref{extent} and Figures~\ref{fig:compact} and \ref{fig:extended} for a description and examples of spatially compact and spatially extended emission. The distribution of spatial separations indicates that the double-peaked AGN emission lines are produced by $\sim$kpc-scale dual AGNs and/or $\sim$kpc-scale outflows, jets, or rotating gaseous disks}
\label{fig:kpc_hist}
\end{figure}

\subsection*{SDSS J000911.58-003654.7} 

For observations of this galaxy, we purposefully aligned one of the slit position angles ($\theta=67^{\circ}.0$) so that it would include the nearby spiral galaxy to the southwest of the target (Figure~\ref{fig:unusual_images}, middle).  This spiral galaxy has photometric redshifts from SDSS of $z_\mathrm{photo}=0.083 \pm 0.017$ (measured with the template fitting method) and $z_\mathrm{photo}=0.066 \pm 0.022$ (measured with a Neural Network method), which are similar to the target galaxy's spectroscopic redshift $z=0.073$.  If the spiral galaxy is indeed at a similar redshift as the target galaxy, then the spatially extended emission we observe (Figure~\ref{fig:unusual_spectra}, middle) could be \hbn.  We measured the wavelength of the centroid of the emission (as in Section~\ref{measure}), and if the emission is indeed \hb then this wavelength corresponds to a spectroscopic redshift $z=0.071$ or a velocity offset 600 km s$^{-1}$ blueward of the target galaxy.
This is suggestive of the type of rich environment that we expect for dual AGNs, which are created from galaxy mergers.

\subsection*{SDSS J100145.30+283330.3} 

This galaxy has an interacting companion to the northwest (Figure~\ref{fig:unusual_images}, bottom), and we aligned one of the slit position angles ($\theta=142^{\circ}.3$) to include this companion.  The long-slit spectra show emission features physically linking the two galaxies (Figure~\ref{fig:unusual_spectra}, bottom), possibly indicating an inflow or outflow of gas between the two galaxies.  This could be an indication of the triggering mechanism that funnels gas onto SMBHs during galaxy mergers.

\section{Results and Discussion}
\label{results}

\subsection{All Double-peaked AGNs We Observed Have Double Emission Components with Kiloparsec-scale Separations}

We find that all of the double-peaked AGNs in our sample exhibit two spatially distinct AGN emission components, with projected separations on the sky ranging from hundreds of pc to a few kpc (the full range is 0.2 $h^{-1}_{70}$ kpc $< \Delta x <$ 5.5 $h^{-1}_{70}$ kpc).  The median separation is 1.1 $h^{-1}_{70}$ kpc, and the distributions of separations are similar for the subsamples with spatially compact and spatially extended emission components (Figure~\ref{fig:kpc_hist}).  We note that, given the 3$^{\prime\prime}$ diameter of an SDSS fiber, emission components with separations $>3^{\prime\prime}$ (which, for our highest-redshift object, corresponds to  $\Delta x >15$ $h^{-1}_{70}$ kpc) will not fall within a single SDSS fiber.

The range of projected separations we measure indicates that the mechanism or mechanisms producing the double-peaked AGN emission lines are neither a very small scale nor a very large scale effect.  Hence, the data are not explained by the rotation of gas in disks of sizes $\simlt$100 pc, bulk flows of gas on small scales near the galaxy's center, small-scale outflows, or AGN pairs or AGN outflows with $\sim$10 kpc separations.  Instead, the spatial separations we measure indicate that the double-peaked AGN emission lines in our sample are produced by $\sim$kpc-scale dual AGNs and/or $\sim$kpc-scale outflows, jets, or rotating gaseous disks.

We note that \cite{SM12.1} identify 30 double-peaked AGNs with peaks that have symmetric profiles or are nearly equal in flux and suggest that they are the products of rotating gaseous disks.  Six of our objects are included in their sample of 30:  SDSS J081542.53+063522.9, SDSS J084049.47+272704.8, SDSS  J124813.82+362423.6, \\ SDSS  J130724.08+460400.9, SDSS J171544.05+600835.7 (which X-ray observations indicate is a dual AGN system; \citealt{CO11.2}), and SDSS J210449.13-000919.1.

\begin{figure}[!t]
\begin{center}
\includegraphics[height=3.5in]{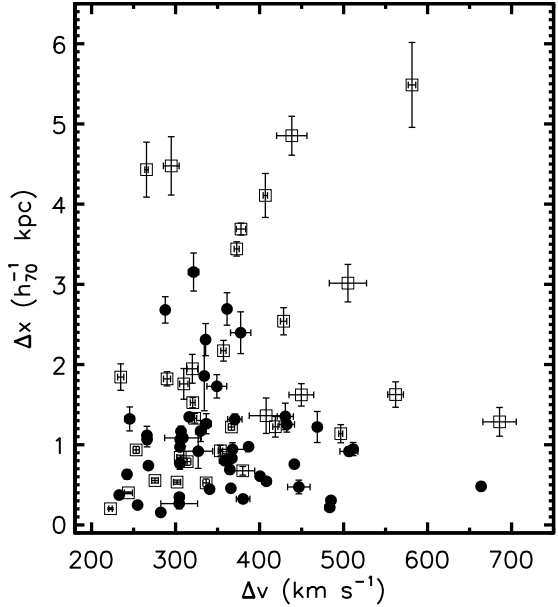}
\end{center}
\caption{Relation between projected physical separations on the sky measured between double \oiiiw emission components in the long-slit spectra and line-of-sight velocity separations between double-peaked \oiiiw emission lines in the SDSS spectra. The objects with spatially extended emission in the long-slit spectra (open squares) have a Spearman's rank correlation coefficient of 0.34 (where the significance level of its deviation from zero is 0.05), suggesting they may be preferentially produced by AGN outflows.  In contrast, the objects with compact emission in the long-slit spectra (filled circles) have no significant correlation, suggesting they may be preferentially produced by dual AGNs.}
\label{fig:kpcv}
\end{figure}

\begin{figure}[!t]
\begin{center}
\includegraphics[height=3.5in]{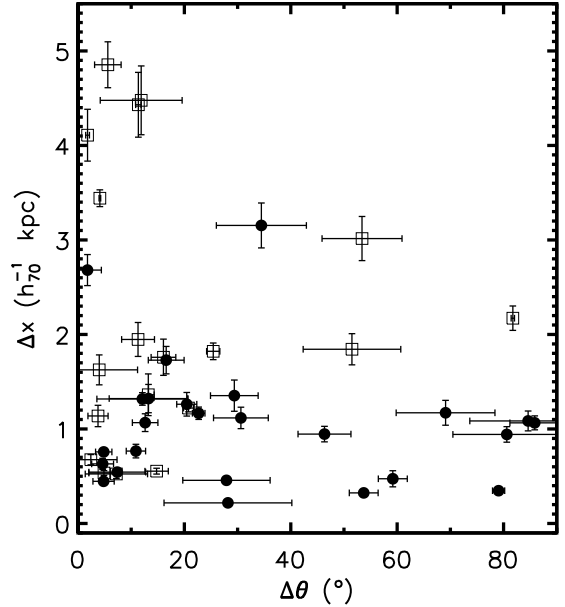}
\end{center}
\caption{Relation between projected physical separations on the sky measured between double \oiiiw emission components in the long-slit spectra and the difference $\Delta\theta$ between the measured position angle of the emission on the sky (from long-slit spectra) and the isophotal position angle of the major axis of the object (from SDSS $r$-band photometry).  Since $\Delta\theta$ is only a meaningful measurement for galaxies with some degree of ellipticity, we plot only the 44 of our 81 galaxies that have ellipticities $e>0.1$.  Of these, we find that $37^{+12}_{-9} \, \%$ of the objects with spatially extended emission in the long-slit spectra (open squares) and $28^{+10}_{-7} \, \%$ of the objects with spatially compact emission in the long-slit spectra (filled circles) have $\Delta\theta$ values that are, within 3$\sigma$, consistent with zero. These fractions are more than twice the expectation for a uniform distribution of $\Delta\theta$ between 0$^\circ$ and 90$^\circ$, indicating that the double AGN emission components are preferentially produced in the plane of the host galaxy.}
\label{fig:kpctheta}
\end{figure}

\subsection{Physical Separation - Velocity Relation}
\label{xvrelation}

The correlations between the projected physical separations and line-of-sight velocity separations of the double-peaked AGNs could help distinguish between different sources of the emission, since dual AGNs may exhibit different correlations than NLR kinematics in single AGN.

If a subset of our sample are dual AGNs, then the spatial separations and velocity offsets we measure are snapshots of the dual AGN dynamics in the stages after the initial galaxy merger but before SMBH coalescence.  For a fixed, three-dimensional physical separation between dual AGNs, we might naively expect a larger (smaller) line-of-sight velocity for a larger (smaller) projected physical separation.  When the orbiting dual AGNs are oriented almost along the line of sight, the projected physical separation will be small and the bulk of the velocity will be transverse, resulting in a small line-of-sight velocity measurement.  In comparison, dual AGNs with a range of three-dimensional physical separations may wash out any relation between projected physical separation and line-of-sight velocity.

A positive correlation between velocity and physical separation could also result from AGN outflows, where a higher- (lower-) velocity outflow enables ionized gas to emit \oiii out to larger (smaller) distances from the central AGN.
In contrast, the opposite correlation should occur for AGN jets.  If a jet is oriented almost along (transverse to) the line of sight, we should measure a large (small) line-of-sight velocity but a small (large) projected physical separation.

For the full sample, we plot the projected physical separations between the double \oiiiw emission components seen in the long-slit spectra against the line-of-sight velocity separations between peaks in the double-peaked \oiiiw emission lines in the SDSS spectra, and we find that the Spearman's rank correlation coefficient is 0.16 and the significance level of its deviation from zero is 0.15 (Figure~\ref{fig:kpcv}). This significance level means that there is a 15$\%$ probability of observing a correlation this strong or stronger in uncorrelated data purely by chance.

However, the correlation strengthens to a Spearman's rank correlation coefficient of 0.34 (significance 0.05) for the subsample with extended spatial emission and essentially disappears for the subsample with spatially compact emission (correlation coefficient -0.04, significance 0.77; see Section~\ref{extent} for a discussion of extended and compact emission).  The physical separation - velocity relation thus seems to separate into two populations: the spatially extended AGN systems and the spatially compact AGN systems.

The spatially extended AGN population's correlation of increasing projected physical separation with increasing line-of-sight velocity is inconsistent with our expectations for AGN jets.  However, the correlation might be explained if the population's spatially extended emission features are preferentially produced by AGN outflows.  The correlation also matches that for dual AGNs at fixed, three-dimensional separations, but NLR effects would be necessary to explain the spatially extended emission.

The spatially compact AGN population's lack of correlation between projected physical separation and line-of-sight velocity could be explained by dual AGNs with a variety of three-dimensional physical separations.  The spatially compact emission is consistent with the scenario of two compact AGNs with no extended NLR features.

Although the physical separation - velocity relations hint that the spatially extended AGNs may be preferentially produced by outflowing AGNs and the spatially compact AGNs may be preferentially produced by dual AGNs with a range of three-dimensional physical separations, we stress that additional observations are required for confirmation of dual AGNs and outflowing AGNs.  Further, the subpopulations of double-peaked AGNs with spatially compact or spatially extended emission are likely not clearly defined as the products of either dual AGNs or NLR kinematics, as the emission profile of a double-peaked AGN could be produced by a combination of both effects.

\subsection{Double Emission Components Are Preferentially Aligned with the Host Galaxy Major Axis}
\label{align}

Our measurements of the position angles of the double emission components can be used to determine their orientations in the host galaxies.  We define $\Delta\theta$ as the difference between the isophotal position angle of the major axis of the galaxy (which is determined from SDSS $r$-band photometry) and the position angle of the two AGN emission components on the sky (which we measure as $\theta_{\mathrm{sky}}$ in Section~\ref{measure}).  However, the isophotal position angle of the major axis of the galaxy, and hence $\Delta\theta$, is only a meaningful measurement in systems where the photometry shows the galaxy has some degree of ellipticity.  Consequently, we focus only on the 44 of the 81 galaxies in our sample that have ellipticities $e > 0.1$ (where ellipticities are computed in Section~\ref{ellipticities}).

For these 44 galaxies, we find that the position angle of the two AGN emission components on the sky is often aligned with the isophotal position angle of the major axis of the object, such that the difference $\Delta\theta$ is close to zero (Figure~\ref{fig:kpctheta}). In fact, $32^{+8}_{-6} \, \%$ (14 objects) of the full sample of 44 galaxies, $37^{+12}_{-9} \, \%$ (7 objects) of the subset with spatially extended emission, and $28^{+10}_{-7} \, \%$ (7 objects) of the subset with spatially compact emission have $\Delta\theta$ values that are within a 3$\sigma$ deviation from zero.  These fractions are more than twice the expectation for a uniform distribution of $\Delta\theta$: if the errors on $\Delta\theta$ remained the same but the $\Delta\theta$ values were distributed uniformly between 0$^\circ$ and 90$^\circ$, only $14^{+6}_{-4} \, \%$ of the sample would have $\Delta\theta$ values within a 3$\sigma$ deviation from zero.  

Expressed another way, $30^{+7}_{-6} \, \%$ of the 44 galaxies have $\Delta\theta<10^\circ$, which is almost three times the expectation of $11^{+7}_{-3} \, \%$ for a flat distribution of $\Delta\theta$ between 0$^\circ$ and 90$^\circ$.  In addition, $55^{+7}_{-8} \, \%$ of the 44 galaxies have $\Delta\theta<20^\circ$, which is more than twice the expectation of $22^{+8}_{-5} \, \%$ for a flat distribution of $\Delta\theta$ between 0$^\circ$ and 90$^\circ$. 

Alignment of the two AGN emission components with the major axis of the object (or low $\Delta\theta$) indicates that the emission is produced in the plane of the galaxy, which is consistent with emission produced by dual AGNs (whose orbits are driven by the potential of the host galaxy and dynamical friction from the host galaxy stars; in simulations, \citealt{BL12.1} also see alignment of the dual AGNs with the ellipticities of the host galaxies) and some AGN-driven outflows and jets (see, e.g., \citealt{WH04.1, RO10.1}).  Other AGN-driven outflows and jets can be more orthogonal to the host galaxy plane, approaching $\Delta\theta \sim90^\circ$ (e.g., \citealt{WH88.1,CE01.1}), but we do not anticipate many dual AGNs at these orientations.  Based on these expectations, low $\Delta\theta$ could be used as a selector for double-peaked AGNs that are, with additional follow-up observations, more likely to yield detections of dual AGNs than double-peaked AGNs with high $\Delta\theta$.

We also plot $\Delta\theta$ against the line-of-sight velocity separations between \oiiiw emission components, but we find no significant correlation between them.

\subsection{The Most Promising Dual AGN Candidates}
\label{promising}

While all double-peaked AGNs can be considered candidate dual AGNs, we attempt to identify observational signatures that select the most promising candidates.  We find that two spatially compact AGN emission components, both emission components in the plane of the host galaxy, and two stellar nuclei spatially coincident with the two emission components may be indicators of dual AGNs.
In this way, we can focus limited observational resources for follow-up on the double-peaked AGNs that are more likely to yield dual AGN detections. We report these promising candidates in Sections~\ref{plane} and~\ref{ao}, but definitive proof of dual AGNs requires the spatial resolution of X-ray or radio cores associated with each AGN (for details see \citealt{CO11.2,FU11.3,TI11.1}).  

\subsubsection{Fourteen Double-peaked AGNs with Emission in the Plane of the Host Galaxy}
\label{plane}

Since the dynamics of kpc-scale separation dual AGNs are dominated by the host galaxy potential and dynamical friction from the host galaxy's stars, dual AGNs should preferentially reside in the plane of the host galaxy.
In Section~\ref{align}, we used this expectation to identify 14 promising dual AGN candidates that have double optical AGN emission components oriented in the plane of the host galaxy, where the difference in position angles $\Delta\theta$ are within 3$\sigma$ deviations from zero.  These 14 candidates include seven double-peaked AGNs with spatially extended emission (SDSS J084049.47+272704.8,  SDSS J095207.62+255257.2, SDSS J160524.59+152233.5, SDSS J163316.03+262716.3, SDSS J210449.13-000919.1, SDSS J225510.12-081234.4, SDSS J230442.82-093345.3) and seven double-peaked AGNs with spatially compact emission (SDSS J011659.59-102539.1, SDSS J080418.23+305157.2, SDSS J093024.84+343057.3, SDSS J102325.57+324348.4, SDSS J144804.17+182537.9, SDSS J162939.58+240856.0, SDSS J225420.99-005134.1).  The seven double-peaked AGNs with spatially compact emission may be the most promising dual AGN candidates of the 14, since we showed in Section~\ref{xvrelation} that spatially compact systems may be preferentially produced by dual AGNs.

\subsubsection{Three Double-peaked AGNs with Double Stellar and AGN Emission Components that Are Potentially Coincident} 
\label{ao}

\begin{figure}[!t]
\centering
\subfigure{\includegraphics[height=4.6cm]{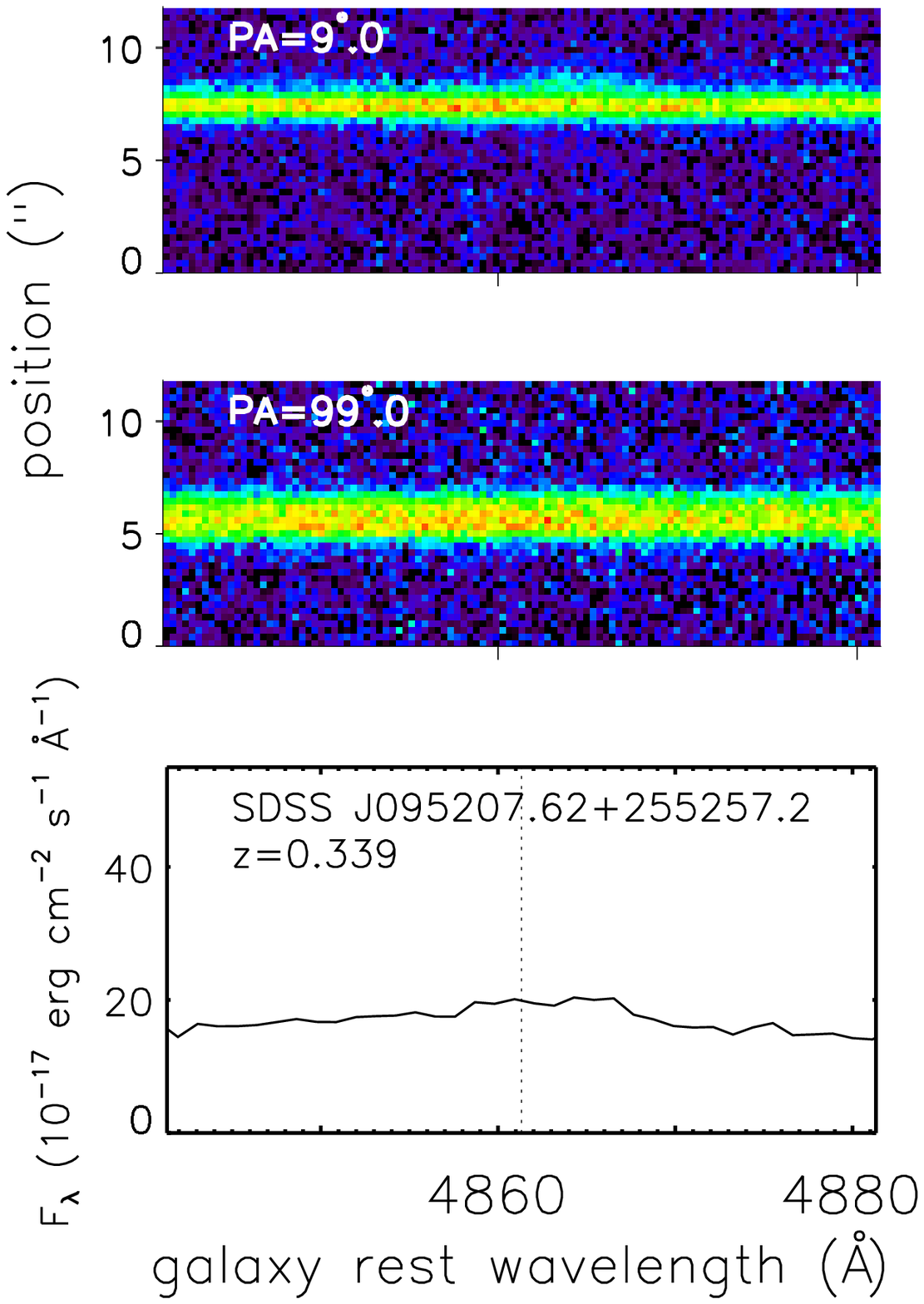}}
\hspace{-0.7cm}
\subfigure{\includegraphics[height=4.6cm]{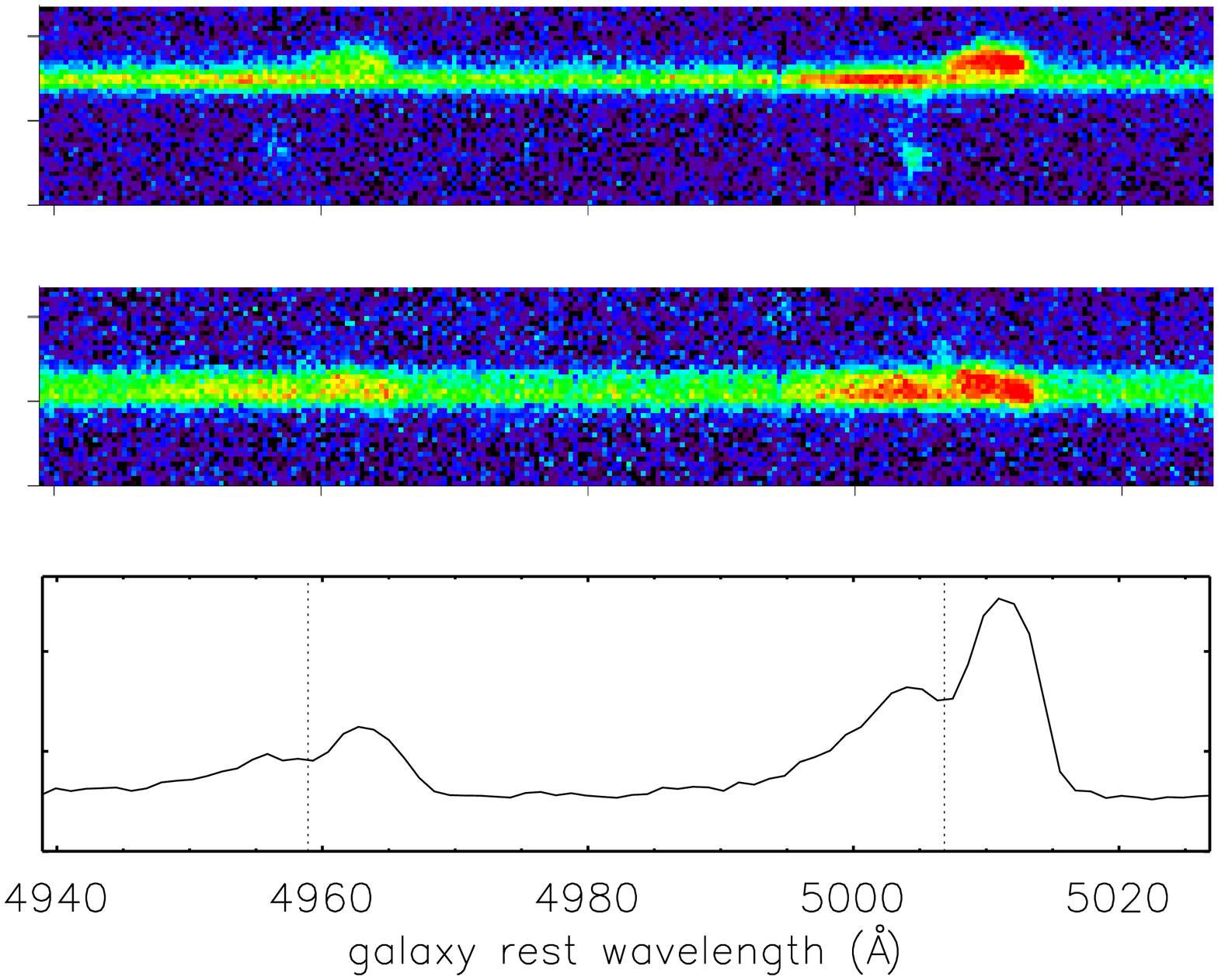}} \\
\subfigure{\includegraphics[height=3.65cm]{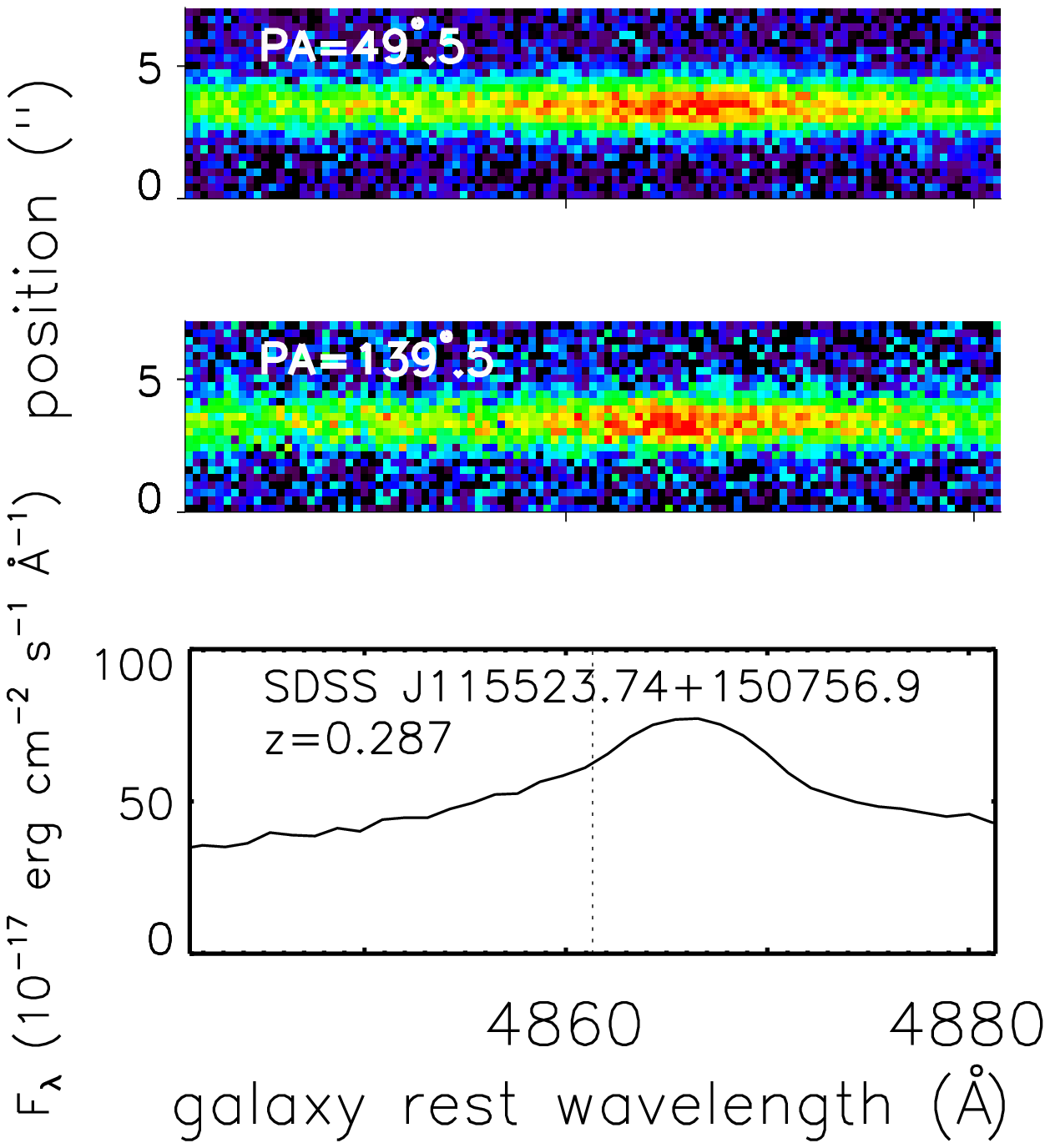}} 
\hspace{-0.7cm}
\subfigure{\includegraphics[height=3.65cm]{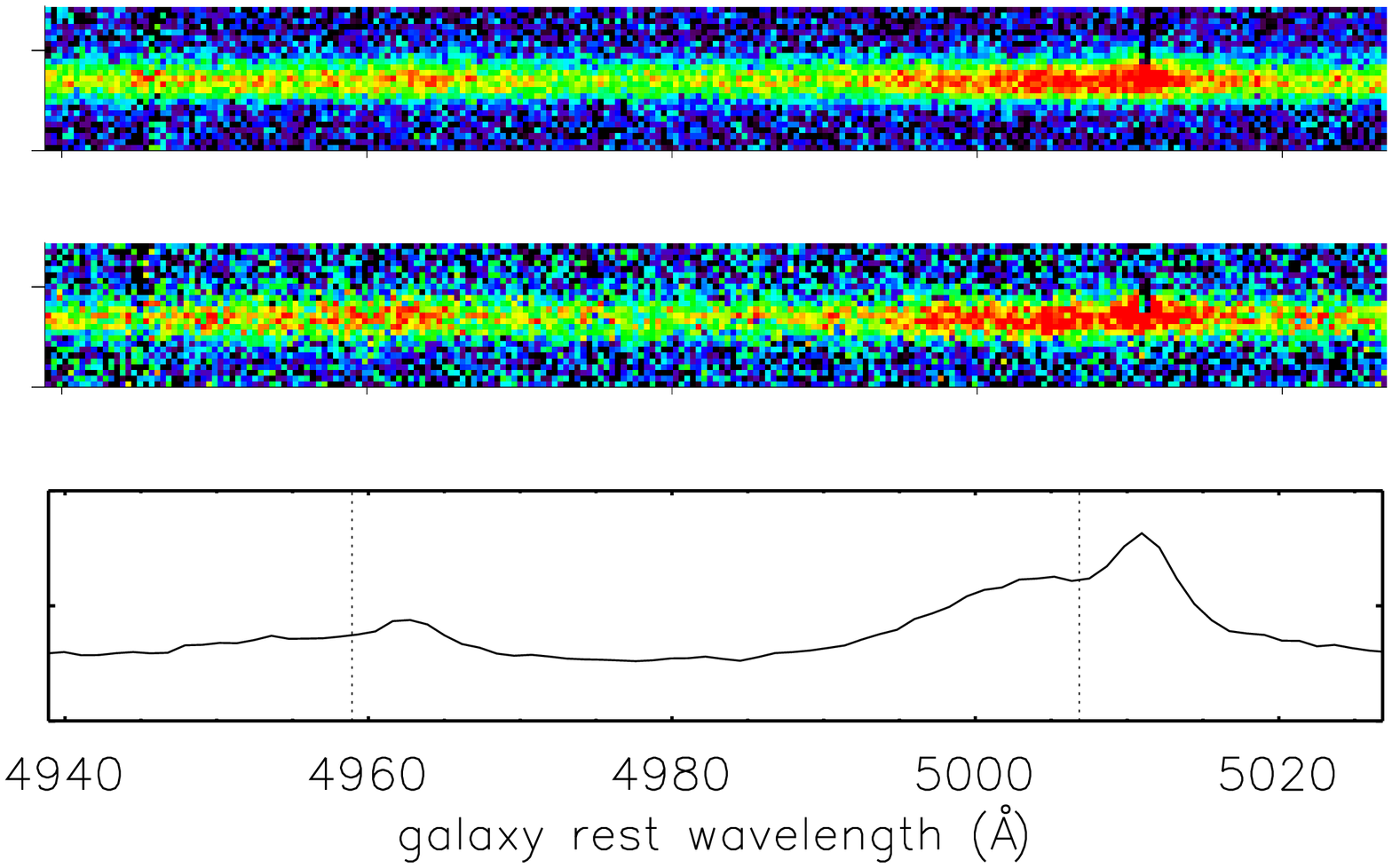}} \\
\subfigure{\includegraphics[height=4.9cm]{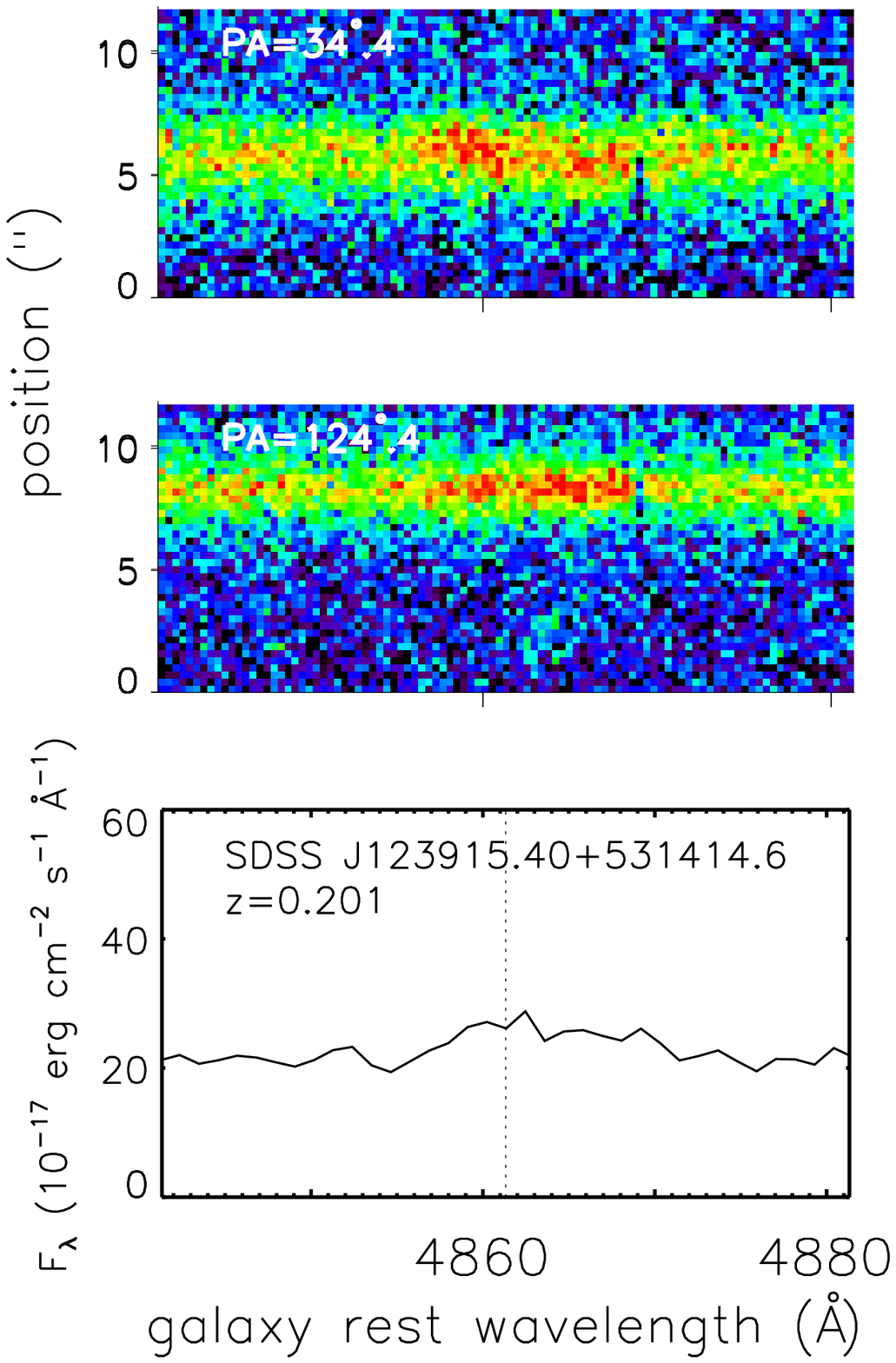}} 
\hspace{-0.7cm}
\subfigure{\includegraphics[height=4.9cm]{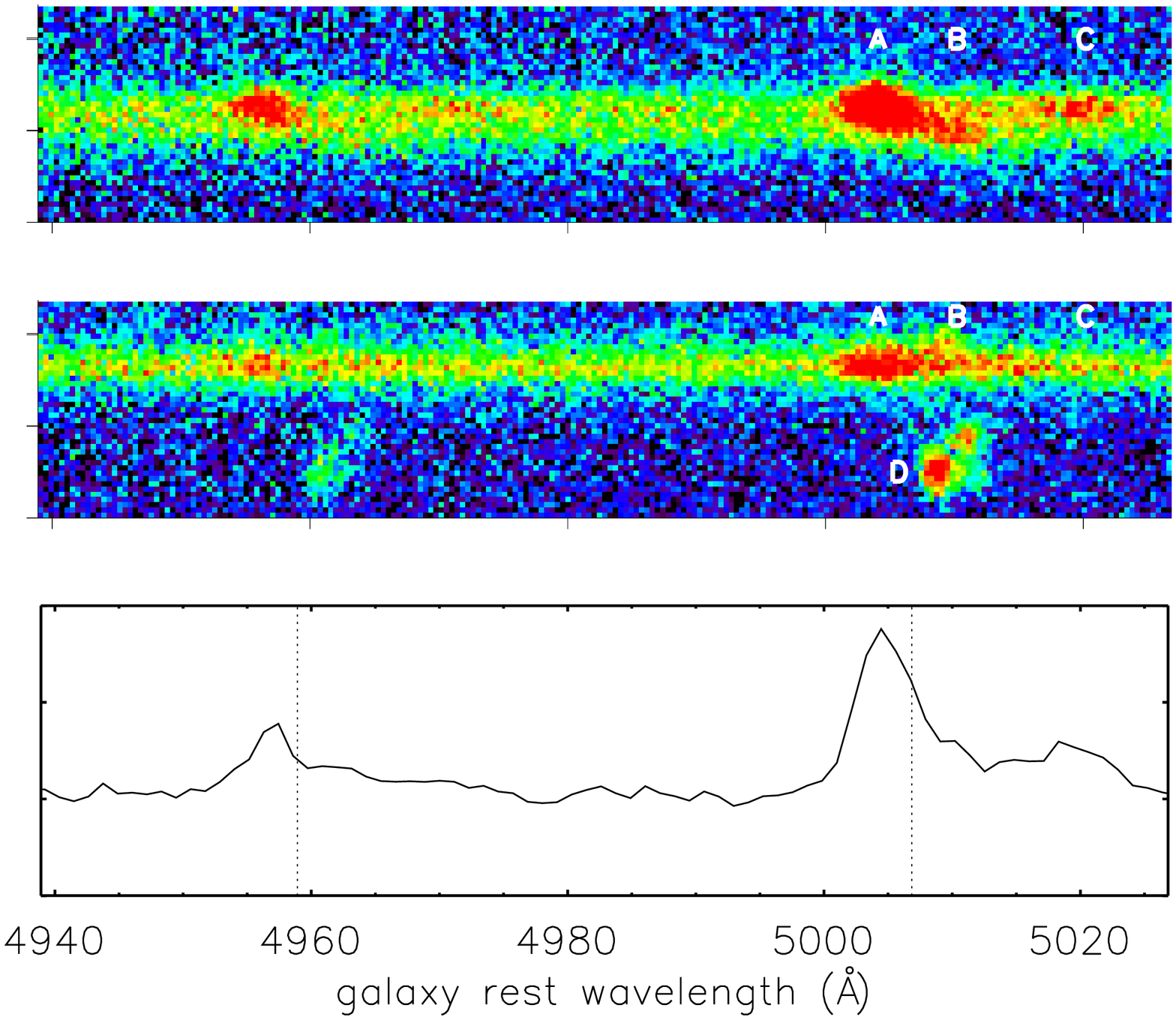}} 
\caption{Same as Figure~\ref{fig:compact}, but for the three double-peaked AGNs that are some of the best dual AGN candidates.  They are promising candidates because the spatial offsets and position angles of their emission features are consistent with those of the double stellar components visible in AO imaging. The A, B, C, and D labels in the bottom long-slit spectra show the positions of the four emission components found in SDSS J123915.40+531414.6 (see Section~\ref{ao} for details).}
\label{fig:dual_candidates}
\end{figure}

Of the seven double-peaked AGN in our sample that exhibit double stellar components in AO imaging (Section~\ref{multiwavelength}), we find that three (SDSS J095207.62+255257.2, SDSS J115523.74+150756.9, and SDSS J123915.40+531414.6) have double AGN emission components with spatial separations and position angles comparable to the spatial separations and position angles between the double stellar components in the corresponding AO image (Figure~\ref{fig:dual_candidates}).  In each case, the spatial separations are consistent to within our 1$\sigma$ errors and the position angles are consistent to $\sim15^\circ$, suggesting that the double AGN emission components are likely coincident with the double stellar components.  This scenario is a signature of a late-stage galaxy merger in which the stellar remnants of the original progenitor galaxies accompany the two AGNs as they evolve into a kpc-scale pair.

Consequently, these galaxies are three of the most compelling dual AGN candidates in our sample.  One of these, SDSS J095207.62+255257.2, was also identified as a dual AGN candidate by \cite{MC11.1} and \cite{FU12.1}, who both deem it a confirmed dual AGN system based on integral-field spectroscopy that shows AGN emission coincident with double stellar components in the AO imaging. While double stellar nuclei coincident with double optical or near-infrared AGN emission components is strong evidence for dual AGNs, such a configuration could also be explained by a single AGN between the two continuum sources.  In addition, dual AGNs may not always have double stellar components visible in AO, as is the case for two systems that are strong dual AGN candidates based on X-ray and other observations but exhibit only a single source in AO imaging \citep{CO11.2,BA12.1}.  In cases like these, the second stellar component could be hidden by dust obscuration or could be too faint to be within the detection threshold.  While the three galaxies discussed here are good candidates for dual AGNs, additional follow-up observations are required to confirm or refute them as dual AGNs.

We note that SDSS J123915.40+531414.6 is an unusual case because the two emission features corresponding to the double peaks in the SDSS spectrum have a significantly smaller spatial offset (1.3 $h^{-1}_{70}$ kpc, or 0$\farcs$39) than the spatial offset between the two stellar components visible in the AO image (4.1 $h^{-1}_{70}$ kpc, or 1$\farcs$24; \citealt{FU11.1}).  Instead, we find a third emission component (labeled B in Figure~\ref{fig:dual_candidates}, bottom) between the blueward (labeled A) and redward (labeled C) features corresponding to the double peaks.  This third emission component has a spatial offset ($4.17 \pm 0.72$ $h^{-1}_{70}$ kpc, or $1\farcs20 \pm 0\farcs21$) and position angle ($32^\circ.5 \pm 3^\circ.3$) relative to the blueward component that is consistent with those of the double stellar components.  Note that there is also a fourth emission component (labeled D), visible only in the two-dimensional spectrum taken at position angle $\theta=124^\circ.4$, that is $\sim5^{\prime\prime}$ from the target galaxy's stellar continuum and may be reflective of an outflow.   The long-slit and AO observations of SDSS J123915.40+531414.6 suggest it might host a triple AGN system or a dual AGN system with outflows.

\section{Conclusions}

We have obtained follow-up long-slit spectroscopy of 81 double-peaked narrow-line AGNs in SDSS at $0.03 \leq z \leq 0.36$ with Lick, Palomar, and MMT Observatories.  The purpose of these observations was to obtain spatial information about the sources of the AGN emission, to aid in distinguishing whether the double-peak morphologies are a result of dual AGNs, rotating gas in a disk, or AGN jets and outflows and on what physical size scales.  We measure the projected spatial separation and position angle of the double AGN emission components on the sky, and compare to the orientation of the host galaxy in SDSS photometry as well as multiwavelength observations of the double-peaked AGNs.  Our main results are summarized below.

\vspace{.1in}

1. In all of the 81 double-peaked AGNs we observe, the long-slit spectra reveal double AGN emission components with projected separations on the sky of 0.2 $h^{-1}_{70}$ kpc $< \Delta x <$ 5.5 $h^{-1}_{70}$ kpc (median $\Delta x = 1.1$ $h^{-1}_{70}$ kpc; Figure~\ref{fig:kpc_hist}). These results indicate that the double-peaked AGN emission lines in our sample are produced neither by very small-scale gas rotation and outflows nor by $\sim$10 kpc scale AGN pairs or outflows. Instead, our observations are consistent with the double AGN emission components originating in $\sim$kpc-scale dual AGNs or $\sim$kpc-scale outflows, jets, or rotating gaseous disks.

\vspace{.1in}

2.  The double-peaked AGNs can be divided into two populations: those with double spatially extended emission components whose line-of-sight velocity differences increase with projected physical separation on the sky (34/81, or $42^{+6}_{-5} \,\%$ of our sample) and those with double spatially compact emission components whose line-of-sight velocity differences are independent of projected physical separation (47/81, or $58^{+5}_{-6} \, \%$ of our sample).  
The physical separation - velocity trends shown in Figure~\ref{fig:kpcv} suggest that the spatially extended sample may be preferentially produced by AGN outflows (since higher-velocity outflows can produce \oiii emission at larger distances) and the spatially compact sample may be preferentially produced by dual AGNs with a range of actual three-dimensional separations. Consequently, we suggest that double-peaked AGNs with double, spatially compact emission components may be fruitful targets for dual AGN searches.

\vspace{.1in}

3.  In $32^{+8}_{-6} \, \%$ of the objects in our sample, we find that the orientations of the double AGN emission components on the sky align with the major axis of the host galaxy, such that the differences $\Delta\theta$ are within a 3$\sigma$ deviation from zero (Figure~\ref{fig:kpctheta}).  This result is more than double the expectation ($14^{+6}_{-4} \, \%$) for a uniform distribution of $\Delta\theta$. Since dual AGNs should orbit in the plane of the galaxy, small $\Delta\theta$ could be a useful indicator for promising dual AGN candidates.  Our proposal of small $\Delta\theta$ as an indicator for dual AGNs is supported by simulations that show a strong alignment between the orientations of dual AGNs and the major axis of the host galaxy \citep{BL12.1}.

\vspace{.1in}

4. We identify 17 particularly promising dual AGN candidates for follow-up observations.  Fourteen double-peaked AGNs (listed in Section~\ref{plane}) have double AGN emission components in the long-slit spectra that are oriented in the plane of their host galaxies, as expected for dual AGNs.  In an additional three double-peaked AGNs (listed in Section~\ref{ao} and shown in Figure~\ref{fig:dual_candidates}), the spatial offset and position angle we measure between the double AGN emission components are consistent with the spatial offset and position angle between the double stellar components in AO imaging.  This suggests that the double AGN emission components may be coincident with the double stellar components, which is a scenario expected for dual AGNs.  

\vspace{.1in}

While the one-dimensional spectra of double-peaked AGNs alone do not reveal much about the sources of the emission, we have shown that the spatial information about the emission provided by follow-up long-slit observations can help distinguish between between gas rotation, AGN outflows and jets, and dual AGNs on different physical scales.  In this way, long-slit observations enable identification of the most compelling dual AGN candidates for confirmation via X-ray or radio observations that spatially resolve two AGNs.

\acknowledgements We thank Jeffrey Silverman and Maryam Modjaz for valuable assistance in learning to reduce the Lick and MMT data, respectively, and Grant Williams for observations during MMT engineering time.  J.M.C. is supported by an NSF Astronomy and Astrophysics Postdoctoral Fellowship under award AST-1102525.  The Texas Cosmology Center is supported by the College of Natural Sciences and the Department of Astronomy at the University of Texas at Austin and the McDonald Observatory.  M.C.C. acknowledges support for this work provided by NASA through Hubble Fellowship grant HF-51269.01-A, awarded by the Space Telescope Science Institute, which is operated by the Association of Universities for Research in Astronomy, Inc., for NASA, under contract NAS 5-26555. M.C.C. also acknowledges support from the Southern California Center for Galaxy Evolution, a multi-campus research program funded by the University of California Office of Research. The observations reported here were obtained at Lick Observatory, a multi-campus research unit of the University of California; the Hale Telescope, Palomar Observatory as part of a continuing collaboration between the California Institute of Technology, NASA/JPL, and Cornell University; and the MMT Observatory, a joint facility of the University of Arizona and the Smithsonian Institution.

{\it Facilities:} \facility{Shane (Kast Double spectrograph)}, \facility{Hale (Double spectrograph)}, \facility{MMT (Blue Channel spectrograph)}

\bibliographystyle{apj}

\end{document}